%% file: main.tex
\documentclass[times, twocolumn]{aastex631}
\usepackage{amsfonts,amsmath,amssymb,amsthm,epsfig,graphicx,float,tabularx,multirow,booktabs,gensymb}
\usepackage{booktabs,color,textcomp}
\usepackage{natbib}
\usepackage{rotating}
\usepackage{makecell}
\usepackage{multirow}
\usepackage{appendix}
\usepackage{color,colortbl}
\usepackage{longtable}
\usepackage[dvipsnames]{xcolor}
\setcellgapes{0pt}

\newcolumntype{S}{>{\centering\arraybackslash}m{0.05\linewidth}}
\newcolumntype{E}{>{\centering\arraybackslash}m{0.05\linewidth}}
\newcolumntype{M}{>{\centering\arraybackslash}m{0.1\linewidth}}

\newcolumntype{N}{>{\centering\arraybackslash}m{0.0005\linewidth}}
\received{September 29, 2025}
\revised{October 28, 2025}
\accepted{November 4, 2025}

\usepackage{xcolor}
\usepackage{amsmath}

\usepackage{graphicx}
\usepackage{natbib}

\usepackage{multirow}
\usepackage{booktabs}
\usepackage{tabularx}

\usepackage{savesym}
\savesymbol{tablenum}
\usepackage{siunitx}
\restoresymbol{SIX}{tablenum}

\usepackage{xspace} % handle spacing after newcommand macros

\begin{document}

\title{CoronaGraph Instrument Reference stars for Exoplanets (CorGI-REx) I. Preliminary Vetting and Implications for the Roman Coronagraph and Habitable Worlds Observatory}

\correspondingauthor{Justin Hom}
\email{jrhom@arizona.edu}

\input{rexcorgi_authors.txt}

\begin{abstract}
The upcoming Roman Coronagraph will be the first high-contrast instrument in space capable of high-order wavefront sensing and control technologies, a critical technology demonstration for the proposed Habitable Worlds Observatory (HWO) that aims to directly image and characterize habitable exoEarths. 
The nominal Roman Coronagraph observing plan involves alternating observations of a science target and a bright, nearby reference star. High contrast is achieved using wavefront sensing and control, also known as ``digging a dark hole”, where performance depends on the properties of the reference star, requiring $V<3$, a resolved stellar diameter $<$2 mas, and no stellar multiplicity. The imposed brightness and diameter criteria limit the sample of reference star candidates to high-mass main sequence and post-main sequence objects, where multiplicity rates are high. A future HWO coronagraph may have similarly restrictive criteria in reference star selection. From an exhaustive literature review of 95 stars, we identify an initial list of 40 primary and 18 reserve reference star candidates relevant to both the Roman Coronagraph and HWO. We present results from an initial survey of these candidates with high-resolution adaptive optics imaging and speckle interferometry and identify no new companions. We discuss the need for higher-contrast observations to sufficiently vet these reference star candidates prior to Roman Coronagraph observations along with the implications of reference star criteria on observation planning for Roman and HWO.

\end{abstract}


\section{Introduction}
The direct detection and characterization of Earth-like, habitable exoplanets around our nearest stellar neighbors is one of the primary science goals for astronomy and astrophysics over the next several decades \citep{decadal2020}. To achieve such an ambitious goal, the \textit{Astro2020} decadal survey proposed development of a large UVOIR space telescope mission, labeled the Habitable Worlds Observatory (HWO), which would possess high-contrast imaging and spectroscopic capabilities for detecting and characterizing nearby exoEarths. Thus far, current state-of-the-art, high-contrast imaging instrumentation has achieved contrast levels of $\sim10^{-7}$ at near-IR wavelengths at separations less than an arcsecond \citep{follette2023,howell2024}. To successfully detect Earthlike exoplanets around the nearest stars, contrasts of order $\sim10^{-10}$ at visible wavelengths and separations of hundreds of milliarcseconds are likely required, imposing a stringent requirement on the performance of high-contrast instrumentation. The path to achieving such high contrast involves precisely controlling stellar wavefronts and nulling contrast-degrading host star speckles in a process known as wavefront sensing and control \citep[WFSC; see also][]{malbet1995}. Such technologies will be demonstrated for the first time in space with the Coronagraph Instrument onboard the Nancy Grace Roman Space Telescope \citep{mennesson2020}. The Roman Coronagraph has one fully-supported observing mode, Hybrid-Lyot Coronagraph (HLC) Band 1 ($\lambda_C = 575$nm), and several ``best-effort" observing modes utilizing shaped-pupil coronagraphs \citep[SPCs; see also][]{riggs2025} with imaging, spectroscopic, and polarimetric capability. In addition to serving as a critical technology pathfinder for HWO, the Roman Coronagraph is also expected to perform ground-breaking science, including the direct imaging and characterization of Jovian exoplanets in reflected light \citep{bailey2023} and circumstellar debris disks in total and polarized intensity \citep[e.g.,][]{anche2023,hom2020}.

The process for high-order wavefront sensing and control (HOWFSC), also known as ``digging a dark hole," requires the use of deformable mirrors to precisely control incoming wavefronts. The efficiency of this approach correlates with the brightness of the star used for wavefront sensing. However, many science targets of interest are not necessarily the brightest stars. As a result, the procedure for a typical Roman Coronagraph observation first involves observing a bright reference star for HOWFSC. Once the dark hole region reaches an appropriate contrast level ($\lesssim10^{-7}$), the deformable mirrors are frozen\footnote{A low-order Zernike control loop still operates but only for low-order wavefront sensing and control \citep{shi2016,shi2018,seo2025}.} and the observatory is slewed to the science target for the acquisition of science images. The contrast of the dark hole, however, will degrade due to drifts and other effects. As a result, the observatory is moved back to the reference star for HOWFSC ``touch-up" and the process repeats \citep[e.g.,][]{cady2025,greenbaum2025real}.

To reach deeper contrasts than $10^{-7}$, e.g. for detecting the reflected light from Jovian exoplanets at $\sim10^{-9}$ contrast, additional post-processing is required, such as through principal component analysis \citep[PCA; e.g.,][]{soummer2012,amara2012}. PCA approaches rely on a set of reference images from which quasistatic speckles arising from the instrumental PSF are well-matched to science images. A library of references can be constructed from science images through angular differential imaging \citep[ADI;][]{marois2006} or observations of another reference star \citep[reference differential imaging; RDI; e.g.,][]{debes2019}. Space telescopes are limited to only a few orientations of spacecraft rolls, with Roman only having a 30$^{\circ}$ range in roll capability. This limits ADI efficacy at small separations (e.g., the inner working angle) and is also detrimental to the imaging of extended objects such as circumstellar disks due to self-subtraction \citep{milli2012}. Therefore, RDI is the assumed standard approach for Roman Coronagraph post-processing and utilizes images of the dark hole dug around the HOWFSC reference star.

Not every bright star in the sky can be used as a reference, as the performances of both HOWFSC and RDI are correlated with both astrophysical and observational properties of the reference star itself. The criteria that define the sample of usable reference stars for the Roman Coronagraph are restrictive, meaning that only a small subset of the 200 brightest stars in the sky can be considered \citep{wolff2024}. Observability constraints of the reference stars could place a strain on scheduling if the number of usable reference stars is small.

The use of an unsuitable reference star for an observational sequence could potentially have negative impacts, such as limiting the achievable raw and post-processed dark hole contrast, increasing the time for dark hole digging to reach acceptable contrast, and/or potentially causing dark hole digging to fail outright. The curation of a full list of usable reference stars for HOWFSC is therefore of utmost importance to minimize observational risk as much as possible prior to Roman Coronagraph science operations.

The goal of this work is to provide an overview of HOWFSC and RDI reference star criteria, to describe the precursor CoronaGraph Instrument Reference stars for Exoplanets (CorGI-REx) observing campaign, and to discuss the significance of determining reference star viability in the context of the Roman Coronagraph and Habitable Worlds Observatory. In \S \ref{sec:criteria}, we describe the rationale for each of the criteria that define the reference star sample. In \S \ref{sec:target_sample}, we discuss the strategies utilized to identify reference star candidates and the application of suitability ranks to reference star candidates after performing a preliminary vetting with literature and archival sources. In \S \ref{sec:observations}-\ref{sec:results}, we present the results of an initial observing campaign to vet reference star candidates with adaptive optics (AO) imaging and speckle interferometry. In \S \ref{sec:futureobs}, we describe ongoing observational efforts in vetting reference star candidates. In \S \ref{sec:discussion}, we discuss other reference star properties and the impacts they may have on science performance along with the implications of reference star constraints on the Roman Coronagraph and a potential HWO coronagraph. In \S \ref{sec:summary} we summarize our findings and discuss the future work needed to vet the sample of reference star candidates prior to Roman Coronagraph science operations.

\section{Roman Coronagraph Reference Star Criteria} \label{sec:criteria}
The primary criteria \citep{wolff2024,krist2023} for reference stars include: 
\begin{enumerate}
    \item $V<3$,
    \item resolved $V$-band uniform disk diameter $\mathrm{UDD}_{V} < 2$ mas, 
    \item possessing no companions that may inhibit wavefront sensing and control algorithms, and
    \item proximity to a given science target.
\end{enumerate}

The motivation for the $V<3$ requirement comes both from budgeting the time to perform HOWFSC and from the time efficiency of obtaining well-exposed (low noise) reference star images suitable for RDI. For stars fainter than $V=3$, integration times for performing HOWFSC are prohibitively long, as sufficient SNR is needed to measure $\lesssim10^{-7}$ contrast speckles. Given the potential tens to hundreds of hours of integration time needed for the most challenging science targets \citep{bailey2023}, we only consider $V<3$ stars to maximize observational efficiency and to ensure sufficient time for demonstrating $SNR\geq5$ on a point source located 6--9 $\lambda/D$ from a star with $V_{\rm AB} \leq 5$ with a contrast ratio of $\geq10^{-7}$, also known as \textit{Threshold Technical Requirement 5} \citep[TTR5;][]{mennesson2022}, along with exercising the best-effort modes.

The resolved stellar diameter criterion originates from coronagraph masks' sensitivity to finite source diameter effects. The fully-supported HLC mask is very sensitive to low-order aberrations. This sensitivity, in the presence of 0.5 mas rms per axis residual tip-tilt jitter, contributes raw contrast ranging from $\sim10^{-9}$ at 3 $\lambda/D$ separation to $<2\times10^{-10}$ for separations $>5\lambda/D$ \citep{krist2023}. The impact of these jitter levels are roughly equivalent to that of a resolved star with diameter equal to 4$\times$ the magnitude of rms jitter per axis, therefore motivating our upper threshold of 2 mas. It is worth noting, however, that the best-effort SPC masks are more robust to low-order aberrations \citep{krist2023}, and can therefore tolerate larger resolved stellar diameters.

\begin{figure}
    \centering
    \includegraphics[width=\linewidth]{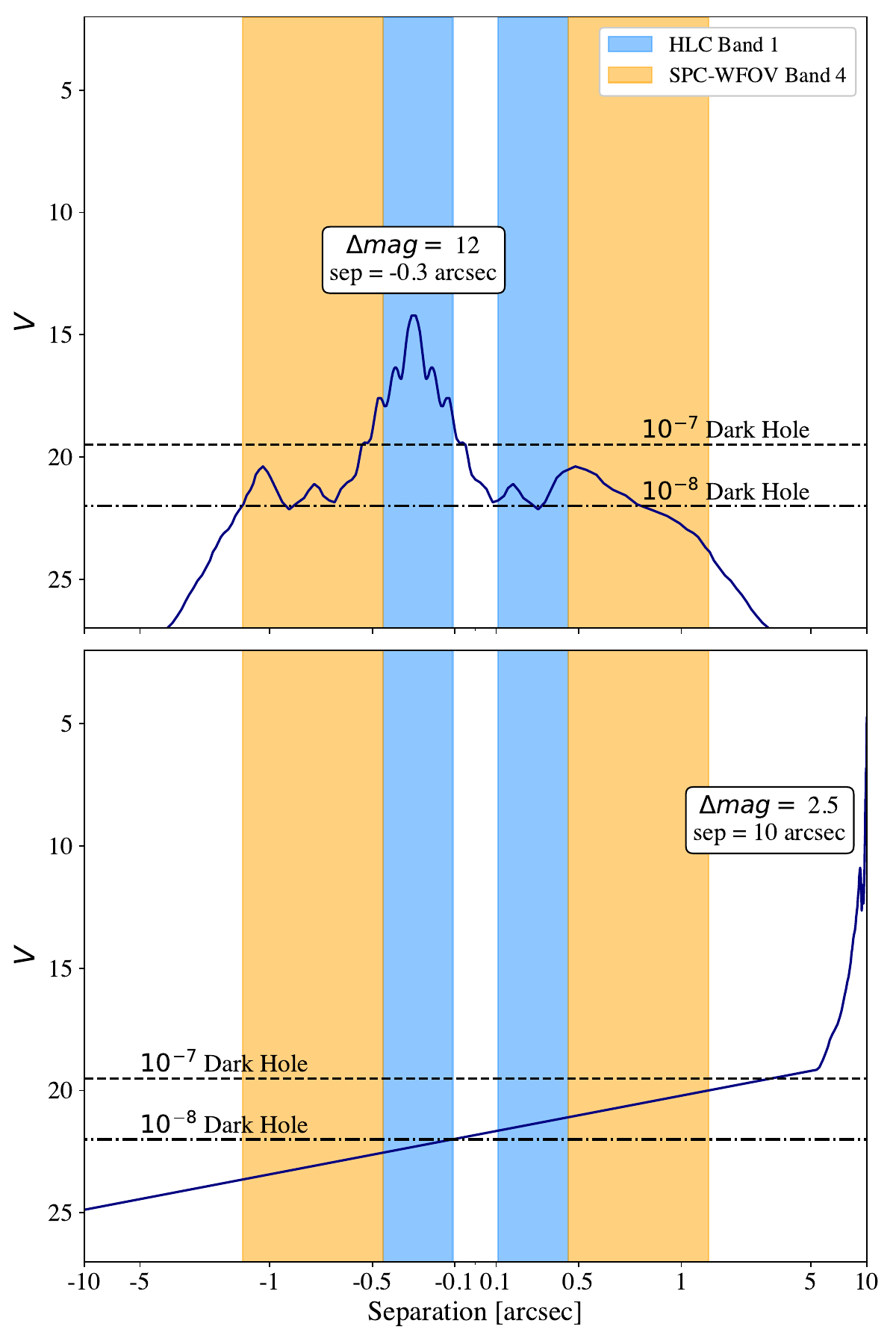}
    \caption{One dimensional representations of how an off-axis companion may leak flux into a Roman Coronagraph dark hole. Both panels assume a $V=2$ reference star, and the navy blue line is the unocculted PSF profile from an off-axis companion at a given $\Delta mag$ and separation. The shaded regions represent the extents of the smallest observing mode FOV (light blue; HLC Band 1) and the largest observing mode FOV (gold; SPC-WFOV Band 4). If sufficient flux from a companion is introduced into the dark hole, both dark hole digging and PSF subtraction via RDI can be compromised.}
    \label{fig:off_axis_glint}
\end{figure}

There are several observing requirements related to the presence of visual reference star off-axis companions for Roman Coronagraph science observations. A Roman Coronagraph observation starts with pointing acquisition on the unocculted reference star in a process known as EXCAM (Exoplanetary Systems Camera) acquisition. EXCAM acquisition requires no sources brighter than 3 magnitudes fainter than the primary target within a $12\arcsec$ radius field-of-view (FOV). Additionally, even companions faint enough to pass EXCAM acquisition may cause issues with HOWFSC or with the utility of the star to serve as a PSF reference for RDI. Interior to the inner working angle, companions cause incoherent light leaks through the coronagraph. Exterior to the outer working angle, the point spread function (PSF) profile of an off-axis companion can introduce incoherent light into the dark hole, degrading the achievable contrast. Within the FOV of each distinct coronagraph observing mode, a reference star should not possess a companion brighter than $10^{-7}$ contrast at the observing wavelength of interest due to impacts on both HOWFSC and RDI suitability. Figure \ref{fig:off_axis_glint} shows two examples of hypothetical $V=2$ stars with off-axis companions at different separations and $\Delta mag$ created from extrapolating models of the unocculted HLC PSF profile from the Roman Coronagraph\footnote{Sourced from \url{https://roman.ipac.caltech.edu/page/additional-coronagraph-instrument-parameters-model-and-data-html}, modeling the amount of scattered light from an off-axis source as a function of separation. Within 10\arcsec, the model is the azimuthally-averaged PSF profile with the focal plane mask and field stop excluded. Values at 45\arcsec and 100\arcsec are estimates of far off-axis scattered light with no coronagraph masks and assume the corresponding PSF profiles are azimuthally-symmetric and achromatic. Overall, the nature of PSF wings at large separations are poorly constrained by a priori calibrations. If using an SPC mask, for example, there will likely be significant azimuthal variation for far off-axis PSF profiles that is not considered for this table.}. In Table \ref{tab:companion_limits}, we summarize the rough minimum $\Delta mag$ thresholds for suitability, assuming a desired raw contrast of $\sim10^{-7}$, for the required HLC Band 1 and best-effort Shaped-Pupil Coronagraph-Wide Field-of-View (SPC-WFOV) Band 4 imaging modes \footnote{The 0.3'' threshold for SPC-WFOV is conservative, as the coronagraph focal plane mask will partially attenuate any companion at this separation.}.

\begin{deluxetable}{ccc}

%% Keep a portrait orientation

%% Over-ride the default font size
%% Use Default (12pt)

%% Use \tablewidth{?pt} to over-ride the default table width.
%% If you are unhappy with the default look at the end of the
%% *.log file to see what the default was set at before adjusting
%% this value.

%% This is the title of the table.
\tablecaption{Estimated minimum allowed $\Delta mag$ for an off-axis companion for two different Roman Coronagraph observing modes and separations for the companion, assuming a desired raw contrast of $\sim10^{-7}$ or better.
} \label{tab:companion_limits}

%% This command over-rides LaTeX's natural table count
%% and replaces it with this number.  LaTeX will increment 
%% all other tables after this table based on this number
%\tablenum{1}

%% The \tablehead gives provides the column headers.  It
%% is currently set up so that the column labels are on the
%% top line and the units surrounded by ()s are in the 
%% bottom line.  You may add more header information by writing
%% another line between these lines. For each column that requries
%% extra information be sure to include a \colhead{text} command
%% and remember to end any extra lines with \\ and include the 
%% correct number of &s.
\tablehead{\colhead{Separation} & \multicolumn2c{Observing Mode} \\ 
\colhead{(arcsec)} & \colhead{HLC Band 1} & \colhead{SPC-WFOV Band 4} } 

%% All data must appear between the \startdata and \enddata commands

\startdata
0.3 & 17.5 & 17.5 \\
0.5 & 15 & 17.5 \\
1.5 & 12 & 14 \\
5 & 4.5 & 6 \\
10 & 3 & 3 \\
20 & 0.5 & 1 \\
\enddata

%% Include any \tablenotetext{key}{text}, \tablerefs{ref list},
%% or \tablecomments{text} between the \enddata and 
%% \end{deluxetable} commands

%% General table comment marker
\tablecomments{Estimates at separations greater than $10\arcsec$ are more uncertain, as they rely on interpolation between single point model predictions for $45\arcsec$ and $100\arcsec$ off-axis.}

%% No \tablerefs indicated

\end{deluxetable}

Finally, reference stars must be within 5$^{\circ}$ of observatory pitch angle from the science target. Assuming that the observatory rolls about the roll vector $\vec{x}$ and slews along perpendicular vectors yaw ($\vec{y}$) and pitch ($\vec{z}$), a changing pitch angle corresponds to a change in the obliquity of solar illumination of the telescope solar panel array in addition to changing geometry of illumination of the spacecraft bus as it slews between targets. Early simulations showed that a pitch angle change (hereafter $\Delta$pitch) of $\geq15^{\circ}$ induces substantial contrast-degrading thermal variations as the observatory slews between reference and science targets \citep{krist2023}. The magnitude of contrast degradation as a function of $\Delta$pitch is not well-known, as the only other simulation performed considered $\Delta$pitch $=3.5^{\circ}$. Little contrast degradation occurred in this simulation, and due to the inherent risk of allowing larger $\Delta$pitch angles, the current Roman Coronagraph operational plan is to only consider reference stars where $|\Delta$pitch$|\leq5^{\circ}$ \citep{wolff2024}. There is no contrast degradation found as a function of changing yaw about the Sun line, meaning that a science-reference pair could be widely separated in yaw as long as $\Delta$pitch does not exceed 5$^{\circ}$ and other pointing constraints are met. The observatory can slew between targets at roughly 3$^{\circ}$ per minute. Slew times, however, are typically less than HOWFSC operational time. While reference-science target pairing will select options that minimize slew times when feasible, the limited number of options and imposed scheduling restrictions inhibit reference star selectiveness, and other criteria, such as reference star magnitude for decreasing HOWFSC operational time, will take precedence.

\section{The CoronaGraph Instrument Reference stars for Exoplanets Imaging Campaign} \label{sec:target_sample}
The CorGI-REx campaign aims to vet potential reference star candidates prior to Roman Coronagraph science operations. Sufficient vetting requires an exhaustive investigation of literature, archival data, and newly obtained data sources.

\subsection{Definition of the Sample} \label{sec:first_vetting}
We created the initial CorGI-REx sample by querying both the Jean-Marie Mariotti Centre (JMMC) Stellar Diameters Catalogue \citep[JSDC;][]{bourges2014,bourges2017} and the JMMC Measured Diameters Catalogue \citep[JMDC;][]{duvert2016,chelli2016} for all objects where $V<3$ and resolved $V$-band uniform disk diameter $\mathrm{UDD}_{V} \leq 2$ mas. In addition to our primary list of reference star candidates, we also create a reserve list where the diameter criterion is expanded to $\mathrm{UDD}_{V} < 5$ mas. This reserve list can be used for the best-effort SPC modes with little to no contrast degradation, and could still be used for the required HLC mode at the cost of some level of contrast degradation \citep{krist2023}. The combination of brightness and small resolved diameter limit the sample of possible reference star candidates to exclusively high-mass stars, including OBA main sequence and BAFGK post-main sequence giants. This is immediately problematic in tandem with the third major reference star criterion on multiplicity, as stellar multiplicity rates around high-mass stars exceed 50\% and may be as high as 90\% \citep{duchene2013}.

For each object returned from the query, we flagged all objects classified in Simbad as double stars, eclipsing binaries, or spectroscopic binaries. When available, we queried the Washington Double Stars Catalog \citep[WDSC;][]{mason2001} and removed any reference star candidates with sufficiently small ($\lesssim2$\arcsec) separation and $\Delta mag$ ($\Delta mag\lesssim 12$) companions. If any candidates were identified as spectroscopic binaries from \textit{SBC9} \citep{pourbaix2004}, they were also removed from consideration. While we also queried these candidates in other spectroscopic binary catalogs \citep{eggleton2008,chini2012} and the Bright Stars Catalog \citep{hoffleit1991}, we did not immediately remove candidates flagged as spectroscopic binaries, as the results (and in some cases the sources of the presented results) were not always well-described and/or were tentative/inconclusive. We then queried \cite{kervella2019}, \cite{kervella2022}, and \cite{zacharias2022} to obtain proper motion anomaly and astrometric binary information for any reference star candidates observed with \textit{Gaia} and/or \textit{Hipparcos}, including the renormalized unit weight error (RUWE) and the tangential velocity anomaly $dVt$. We also queried the JMMC \textit{Bad Calibrators Catalogue for optical interferometry} \citep[BadCal;][]{lafrasse2010}. We then investigated the literature references associated with each remaining candidate, identifying other stellar multiplicity investigations that utilized either radial velocity, astrometry, and/or direct imaging techniques to identify stellar companions. Finally, we queried the Genereal Catalogue of Variable Stars \citep[GCVS;][]{samus2017} and the American Association of Variable Star Observers international database \citep[AAVSO;][]{percy1993aavso}. We flagged any candidates with stellar variability amplitudes $>100$ mmag over dark hole digging timescales, as large magnitude and rapid variability may negatively impact HOWFSC efficiency\footnote{No remaining candidates after excluding stars with confirmed problematic companions had strong and rapid variability above this threshold.}.

This preliminary literature and catalog vetting left a total of 40 reference star candidates in the primary list and an additional 18 reference star candidates in the reserve list. The list of primary and reserve reference star candidate properties including spectral types, Johnson $V$ and $I$ magnitudes, $\mathrm{UDD}_V$, limb-darkened diameter $\mathrm{LDD}$, distance D, Gaia DR3 RUWE, tangential velocity anomaly $dVt$, and age are listed in Tables \ref{tab:rex-corgi-sample-table} and \ref{tab:rex-corgi-reserve-table}. A list of rejected reference star candidates and the rationale for their rejection is given in Appendix \ref{sec:rejected_candidates}.

\subsection{Preliminary Ranking} \label{sec:refstar_ranks}
\begin{figure*}
    \centering
    \includegraphics[width=\linewidth]{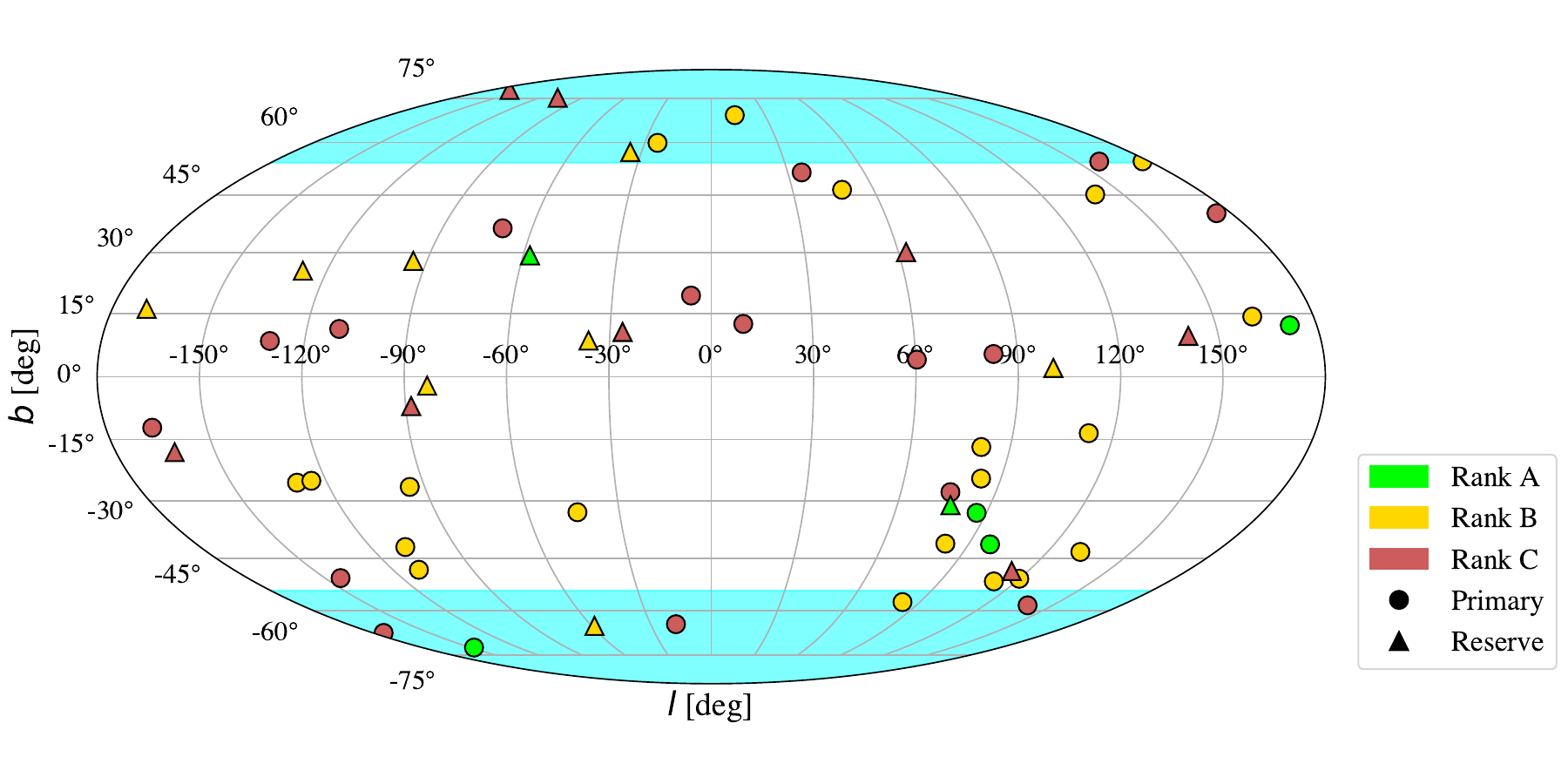}
    \caption{Primary (circles, $\leq2$ mas) and reserve (triangles, $<5$ mas) candidate reference stars plotted in ecliptic coordinates in epoch J2027 and equinox 2027 with colors indicating rankings. The CVZs are shaded in cyan. Only one rank A star is in any of the CVZs, and the overall number of rank A reference star candidates is very low compared to other ranks. If only Rank A stars are viable reference star candidates, science target scheduling for the Roman Coronagraph will likely be restrictive.}
    \label{fig:refstar_sky_coverage}
\end{figure*}

The remaining candidates after preliminary vetting cannot be considered as ``equally suitable," as the reference star candidates have varying degrees of indirect evidence of stellar companions. For the purposes of observation planning for the Roman Coronagraph, we apply a rank of ``A," ``B," or ``C" to each reference star candidate, with the confidence of suitability for Roman Coronagraph observations decreasing from ``A" to ``C." These rankings will ultimately allow for optimized observation planning and contingency plans should some reference stars be identified as unsuitable. 
\begin{itemize}
    \item An A-rank candidate has been investigated through either/both indirect and/or direct methods with no significant evidence of a companion. 
    \item A B-rank candidate has marginal and/or conflicting evidence of a companion from previous indirect and/or direct investigations.
    \item A C-rank companion either has stronger indirect or direct evidence of possessing a companion that has thus far not been confirmed, or has a known companion at sufficiently faint brightness and wide separation to not impact HOWFSC efficiency but may impact RDI efficacy.
\end{itemize}

Ultimately, the preliminary rankings for each reference star candidate are determined quasi-qualitatively, with divisions in rankings loosely correlated to the amount of significant evidence of a companion. This means that stars that have not been extensively characterized (e.g., lacking radial velocity and/or astrometric measurements) are considered rank ``A" along with stars where there is strong evidence against a companion. These preliminary rankings are not considered final because of the high degree of uncertainty in multiplicity from spectroscopic and astrometric analyses. One of the primary challenges in characterizing spectroscopic stellar companions is weaker absorption features in higher-mass stars. Stellar variability and activity indicators also complicate determinations of multiplicity from spectroscopy. Furthermore, any astrometric measurements and inferences originating from \textit{Gaia} photometry are inherently uncertain since all reference star candidates are brighter than the \textit{Gaia} bright limit of $G\sim3$ \citep{lindegren2018}. Despite this greater uncertainty in astrometric measurements, these results are still utilized (albeit at lower impact) in our initial rankings of reference star candidates.

The indirect studies into these stars are blind to background objects, which can only be detected through direct imaging. As problematic background objects are discovered, confidence in current reference star candidates may decrease or be eliminated from consideration entirely. 18 primary and reserve reference star candidates are within $\pm5^{\circ}$ of the galactic midplane, where the likelihood of background contamination increases.

Rankings currently do not consider the presence of bright circumstellar dust that may also degrade achievable contrast through either/both HOWFSC and RDI. $\sim20\%$ of the current list of 40 primary and 18 reserve candidates exhibit an IR excess indicative of cicumstellar dust. While there is a possibility that the dust scattering of circumstellar dust grains may inhibit HOWFSC and/or RDI, we do not exclude or rank these candidates differently in this initial catalog but may consider this factor in future iterations of the reference star candidate list. A detailed discussion of our rationale and the possible impact on performance is given in \S \ref{sec:otherproperties}.

Rankings and science-reference target pair selections also do not consider matching stellar colors between science and reference stars. RDI performance degrades due to chromatic speckle morphologies illuminated by different spectral slopes integrated across the observing bandpass \citep[e.g.,][]{debes2019}, but this degradation is traded favorably against the benefits of bluer reference stars, which are brighter for a given angular diameter. For A-type stars and earlier, the spectral slope across the 575-nm band is similar, so there is little difference in RDI degradation when matching O and G reference and target stars relative to matching A and G stars, for example. This statement is also true for the longer bandpasses used by the Roman Coronagraph. This has been demonstrated in post-processing of simulated datasets \citep{ygouf2025roman}.

Rankings are intended to describe confidence in suitability for Roman Coronagraph observations in general. Specific reference star selection for observing programs, however, may need to follow more discriminating criteria depending on the science goal of a program and/or desired contrast performance. Understanding the nuances of specific reference star selection beyond rankings is left to future work.

A detailed description of the rationale behind each reference star ranking is given in Appendix \ref{sec:rankings_detailed}. The primary and reserve reference star candidates are plotted in ecliptic coordinates in Figure \ref{fig:refstar_sky_coverage}, with colors corresponding to ranks and shapes indicating ``primary" or ``reserve." The continuous viewing zones (CVZs), defined as the regions of the sky where the pointing angle between the observatory and the Sun are between 54$^{\circ}$ and 126$^{\circ}$ \citep{holler2018}, are shaded in cyan; targets located in the continuous viewing zones are always visible from the observatory.

\begin{deluxetable*}{cccccccccccc}

\tabletypesize{\footnotesize}

%% Keep a portrait orientation

%% Over-ride the default font size
%% Use Default (12pt)

%% Use \tablewidth{?pt} to over-ride the default table width.
%% If you are unhappy with the default look at the end of the
%% *.log file to see what the default was set at before adjusting
%% this value.

%% This is the title of the table.
\tablecaption{Primary reference star candidates and properties.} \label{tab:rex-corgi-sample-table}

%% This command over-rides LaTeX's natural table count
%% and replaces it with this number.  LaTeX will increment 
%% all other tables after this table based on this number
%\tablenum{1}

%% The \tablehead gives provides the column headers.  It
%% is currently set up so that the column labels are on the
%% top line and the units surrounded by ()s are in the 
%% bottom line.  You may add more header information by writing
%% another line between these lines. For each column that requries
%% extra information be sure to include a \colhead{text} command
%% and remember to end any extra lines with \\ and include the 
%% correct number of &s.
\tablehead{\colhead{Name} & \colhead{HD} & \colhead{HIP} & \colhead{SpTy} & \colhead{$V$} & \colhead{$I$} & \colhead{$\mathrm{UDD}_V$} & \colhead{$\mathrm{LDD}$} & \colhead{D} & \colhead{Gaia DR3} & \colhead{$dVt$} & \colhead{Age}\\
\colhead{} & \colhead{} & \colhead{} & \colhead{} & \colhead{} & \colhead{} & \colhead{(mas)} & \colhead{(mas)} & \colhead{(pc)} & \colhead{RUWE} & \colhead{(m/s)} & \colhead{(Myr)} 
}

%% All data must appear between the \startdata and \enddata commands
\startdata
\multicolumn{12}{c}{A-Rank Stars}\\\hline
$\kappa$ Ori & 38771 & 27366 & B0.5Ia (1) & 2.06 & 2.27 & 0.44* (2) & ... & 198.41 (3) & 51.25** & ... & $\sim 5$ (4) \\
$\beta$ CMa & 44743 & 30324 & B1II-III (1) & 1.97 & 2.35 & 0.52* (5) & 0.54* (5) & 151.06 (3) & 50.48** & ... & $\sim 13$ (6) \\
$\beta$ Leo & 102647 & 57632 & A3Va (7) & 2.13 (7) & 2.06 & 1.40 & 1.47 & 11.00 (3) & ... & ... & $\sim 50$ (8) \\
$\beta$ Car & 80007 & 45238 & A1III- (9) & 1.69 & 1.61 & 1.60 & 1.68 & 34.70 (3) & ... & ... & $\sim 260$ (10) \\ \hline
\multicolumn{12}{c}{B-Rank Stars}\\\hline
$\epsilon$ Ori & 37128 & 26311 & B0Ia (1) & 1.69 & 1.93 & 0.63* (5) & 0.66* (5) & 360.00 (3) & ... & ... & $\sim 5$ (4) \\
$\delta$ Cas & 8538 & 6686 & A5IV (11) & 2.68 (12) & 2.44 & 1.20 & 1.26 & 30.48 (3) & 4.49 & 66.39$\pm$67.66 & $\sim 600$ (8) \\
$\alpha$ Ara & 158427 & 85792 & B2Vne (13) & 2.95 & 3.29 & 0.67 & 0.68 & 81.97 (3) & 20.30** & 6808.35$\pm$2877.59** & ... \\
$\eta$ Cen & 127972 & 71352 & B2Ve (13) & 2.31 & 2.59 & 0.57 & 0.59 & 93.72 (3) & 4.84** & ... & $\sim 5.6$ (14) \\
$\rho$ Pup & 67523 & 39757 & F5IIkF2IImF5II (15) & 2.81 & 2.25 & 1.50 & 1.59 & 19.46 (16) & 4.38 & 49.07$\pm$26.75 & $\sim 2000$ (8) \\
$\eta$ UMa & 120315 & 67301 & B3V (17) & 1.86 & 2.16 & 0.75 & 0.77 & 31.87 (3) & ... & ... & $\sim 10$ (14) \\
$\gamma$ Ori & 35468 & 25336 & B2III (1) & 1.64 & 1.95 & 0.69 & 0.71 & 77.40 (3) & ... & ... & $\sim 25$ (14) \\
$\alpha$ Cyg & 197345 & 102098 & A2Ia (1) & 1.25 & 1.04 & 2.06 & 2.16 & 432.9 (3) & ... & ... & $\sim 11.6$ (14) \\
$\beta$ Lup & 132058 & 73273 & B2III (18) & 2.68 & 2.95 & 0.47 & 0.48 & 117.37 (3) & ... & ... & $\sim 25$ (14) \\
$\alpha$ Lep & 36673 & 25985 & F0Ib (19) & 2.57 & 2.14 & 1.61 & 1.69 & 680.27 (3) & 2.23** & 959.07$\pm$1304.26 & $\sim 13$ (20) \\
$\delta$ Leo & 97603 & 54872 & A5IV(n) (15) & 2.53 (7) & 2.41 & 1.27 & 1.34 & 17.91 (3) & ... & ... & $\sim 700$ (21) \\
$\beta$ UMa & 95418 & 53910 & A1IVps (22) & 2.37 & 2.35 & 1.09 & 1.14 & 25.90 (16) & 5.99 & 272.13$\pm$123.63 & $\sim 400$ (21) \\
$\eta$ CMa & 58350 & 35904 & B5Ia (1) & 2.45 & 2.44 & 0.82 & 0.84 & 609.76 (3) & ... & ... & $\sim 8$ (14) \\
$\alpha$ Cep & 203280 & 105199 & A8Vn (15) & 2.46 (12) & 2.11 & 1.36 & 1.43 & 15.04 (3) & ... & ... & $\sim 900$ (21) \\
$\gamma$ TrA & 135382 & 74946 & A1V (23) & 2.89 & ... & 1.01 & 1.06 & 58.20 (16) & 2.82 & 186.95$\pm$106.83 & $\sim 100-300$ (21) \\
$\epsilon$ CMa & 52089 & 33579 & B1.5II (1) & 1.50 & 1.80 & 0.77* (2) & 0.80* (2) & 124.22 (3) & ... & ... & $\sim 18$ (24) \\
$\alpha$ Col & 37795 & 26634 & B9Ve (13) & 2.65 & 2.74 & 0.81 & 0.84 & 87.71 (15) & 2.12 & 265.51$\pm$236.82 & $\sim 93$ (13) \\
$\beta$ TrA & 141891 & 77952 & F1V (9) & 2.85 & 2.38 & 1.38 & 1.46 & 12.42 (15) & 2.86 & 63.51$\pm$16.68 & $\sim 675$ (21) \\
$\alpha$ Gru & 209952 & 109268 & B6V (9) & 1.71 & 1.83 & 1.02 & 1.05 & 30.97 (3) & ... & ... & $\sim 100$ (25) \\
$\beta$ CMi & 58715 & 36188 & B8Ve (9) & 2.89 & 2.96 & 0.68 & 0.70 & 49.58 (3) & 27.62** & 1931.04$\pm$549.94** & $\sim 160$ (26) \\ \hline
\multicolumn{12}{c}{C-Rank Stars}\\\hline
$\alpha$ Peg & 218045 & 113963 & A0IV (23) & 2.48 & 2.50 & 0.86 & 0.90 & 40.88 (3) & ... & ... & $\sim 200$ (27) \\
$\zeta$ Pup & 66811 & 39429 & O4I(n)fp (28) & 2.25 & 2.58 & 0.41* (2) & 0.42* (2) & 332.23 (3) & ... & ... & $\sim 3$ (29) \\
$\beta$ Cas & 432 & 746 & F2III (15) & 2.27 & 1.77 & 1.86 & 1.97 & 16.78 (3) & ... & ... & $\sim 1000$ (30) \\
$\zeta$ Oph & 149757 & 81377 & O9.2IVnn (28) & 2.56 & 2.50 & 0.53* (31) & 0.54* (31) & 134.97 (15) & 4.49 & 15325.81$\pm$1815.76** & $\sim 3$ (14) \\
$\delta$ Cru & 106490 & 59747 & B2IV (32) & 2.75 (33) & ... & 0.41 & 0.42 & 139.51 (15) & 4.66 & 800.67$\pm$448.03 & $\sim 18$ (14) \\
$\alpha$ Hyi & 12311 & 9236 & F0IV (9) & 2.84 & 2.39 & 1.57 & 1.66 & 19.40 (34) & 3.26** & 2517.68$\pm$199.42** & $\sim 800$ (35) \\
$\eta$ Tau & 23630 & 17702 & B7III (17) & 2.87 & 2.88 & 0.81 & 0.84 & 123.61 (3) & 1.39** & 3325.43$\pm$1288.79** & $\sim 70$ (36) \\
$\iota$ Car & 80404 & 45556 & A7Ib (11) & 2.26 & 1.82 & 1.77 & 1.87 & 234.74 (3) & 18.56** & 132.45$\pm$23.14** & $\sim 56$ (24) \\
$\beta$ Tau & 35497 & 25428 & B7III (1) & 1.65 & 1.76 & 1.12 & 1.16 & 41.05 (3) & ... & ... & $\sim 100$ (26) \\
$\delta$ Crv & 108767 & 60965 & A0IV(n)kB9 (9) & 2.94 & 3.03 & 0.77 & 0.80 & 26.27 (15) & 3.64 & 103.40$\pm$59.42 & $\sim 3-260$ (37) \\
$\epsilon$ UMa & 112185 & 62956 & A1III-IVpkB9 (15) & 1.77 & 1.82 & 1.46 & 1.53 & 25.31 (3) & 17.62** & 1753.38$\pm$334.95** & $\sim 530$ (36) \\
$\beta$ Eri & 33111 & 23875 & A3IV (15) & 2.79 & 2.57 & 1.10 & 1.16 & 27.59 (15) & 2.68 & 1326.00$\pm$48.30 & $\sim 600$ (21) \\
$\alpha^{2}$ CVn & 112413 & 63125 & A0VpSiEu (23) & 2.88 & 3.05 & 0.71 & 0.74 & 30.56 (15) & 8.42 & 127.61$\pm$86.66 & $\sim 1000$ (21) \\
$\beta$ Lib & 135742 & 74785 & B8Vn (38) & 2.62 & 2.76 & 0.75 & 0.78 & 56.75 (3) & ... & ... & $\sim 80$ (26) \\
$\zeta$ Aql & 177724 & 93747 & A0IV-Vnn (15) & 2.99 & 2.98 & 0.77 & 0.80 & 26.16 (15) & 2.52 & 52.44$\pm$44.53 & $\sim 200$ (21) \\
$\gamma$ Peg & 886 & 1067 & B2IV (1) & 2.84 & 3.13 & 0.37 & 0.38 & 143.94 (15) & 1.97 & 1128.96$\pm$432.08 & $\sim 20$ (14) \\
\enddata

%% Include any \tablenotetext{key}{text}, \tablerefs{ref list},
%% or \tablecomments{text} between the \enddata and 
%% \end{deluxetable} commands

%% General table comment marker
\tablecomments{All $V$ and $I$ measurements are Johnson magnitudes (except for $\delta$ Cru) sourced from \cite{ducati2022_johnson}, except when indicated by a different reference. The $\delta$ Cru $V$ measurement is a Tycho $V_T$ measurement from \cite{hog2000_tycho}. All $UDD_{V}$ and $LDD$ values are sourced from JSDC \citep{bourges2017}, except when denoted by (*). Starred values are direct measurements sourced from JMDC \citep{duvert2016}. RUWE is sourced from Gaia Data Release 3 \citep{gaiadr3} and tangential velocity anomaly ($dVt$) is taken from \cite{kervella2022} except when denoted by (**), for which both values are sourced from \cite{kervella2019}.}

%% General table references marker
\tablerefs{1. \cite{negueruela2024}, 2. \cite{hanbury1974}, 3. \cite{vanleeuwen2007}, 4. \cite{voss2010}, 5. \cite{abeysekara2020}, 6. \cite{fossati2015}, 7. \cite{vanbelle2009}, 8. \cite{rhee2007}, 9. \cite{gray2006}, 10. \cite{gaspar2016}, 11. \cite{gray1989}, 12. \cite{oja_1}, 13. \cite{levenhagen2006}, 14. \cite{tetzlaff2011}, 15. \cite{gray2003}, 16. \cite{gaia2020}, 17. \cite{morgan1973}, 18. \cite{hiltner1969}, 19. \cite{gray2001}, 20. \cite{lyubimkov2010}, 21. \cite{david2015}, 22. \cite{phillips2010}, 23. \cite{abt1995}, 24. \cite{neiner2017}, 25. \cite{su2006}, 26. \cite{janson2011},  27. \cite{derosa2014}, 28. \cite{sota2014}, 29. \cite{howarth2019}, 30. \cite{casagrande2011}, 31. \cite{gordon2018}, 32. \cite{houk1975}, 33. \cite{hog2000_tycho}, 34. \cite{gaia2018}, 35. \cite{lachaume1999}, 36. \cite{brandt2015}, 37. \cite{montesinos2009}, 38. \cite{buscombe1965}.}

\end{deluxetable*}

\begin{deluxetable*}{cccccccccccc}

%% Keep a portrait orientation

%% Over-ride the default font size
%% Use Default (12pt)

%% Use \tablewidth{?pt} to over-ride the default table width.
%% If you are unhappy with the default look at the end of the
%% *.log file to see what the default was set at before adjusting
%% this value.

%% This is the title of the table.
\tablecaption{Reserve reference star candidates and properties.} \label{tab:rex-corgi-reserve-table}

%% This command over-rides LaTeX's natural table count
%% and replaces it with this number.  LaTeX will increment 
%% all other tables after this table based on this number
%\tablenum{2}

%% The \tablehead gives provides the column headers.  It
%% is currently set up so that the column labels are on the
%% top line and the units surrounded by ()s are in the 
%% bottom line.  You may add more header information by writing
%% another line between these lines. For each column that requries
%% extra information be sure to include a \colhead{text} command
%% and remember to end any extra lines with \\ and include the 
%% correct number of &s.
\tablehead{\colhead{Name} & \colhead{HD} & \colhead{HIP} & \colhead{SpTy} & \colhead{$V$} & \colhead{$I$} & \colhead{$UDD_V$} & \colhead{$LDD$} & \colhead{D} & \colhead{Gaia DR3} & \colhead{$dVt$} & \colhead{Age} \\ 
\colhead{} & \colhead{} & \colhead{} & \colhead{} & \colhead{} & \colhead{} & \colhead{(mas)} & \colhead{(mas)} & \colhead{(pc)} & \colhead{RUWE} & \colhead{(m/s)} & \colhead{(Myr)} } 

%% All data must appear between the \startdata and \enddata commands
\startdata
\multicolumn{12}{c}{A-Rank Stars}\\\hline
$\alpha$ Aql & 187642 & 97649 & A7Vn (1) & 0.76 & 0.49 & 3.40 & 3.58 & 5.13 (2) & ... & ... & $\sim 88$ (3) \\
$\beta$ Ori & 34085 & 24436 & B8Ia (4) & 0.13 & 0.15 & 2.62 & 2.72 & 264.55 (2) & ... & ... & $\sim 8$ (5) \\ \hline
\multicolumn{12}{c}{B-Rank Stars}\\\hline
$\beta$ Hyi & 2151 & 2021 & G0V (6) & 2.79 & 1.94 & 2.16 & 2.30 & 7.46 (2) & 3.65 & 21.03$\pm$19.35 & ... \\
$\gamma$ Cyg & 194093 & 100453 & F8Ib (7) & 2.23 & 1.40 & 2.65 & 2.81 & 561.80 (2) & 28.02** & ... & $\sim 12$ (8) \\
$\beta$ Aqr & 204867 & 106278 & G0Ib (9) & 2.89 & 1.84 & 2.38 & 2.53 & 167.43 (10) & 3.08 & 55.81$\pm$230.67 & $\sim 50-100$ (11) \\
$\epsilon$ Gem & 48329 & 32246 & G8Ib (9) & 2.98 & 1.40 & 4.40 & 4.71 & 266.83 (10) & 2.40 & 284.94$\pm$354.07 & $\sim 100$ (12) \\
$\epsilon$ Vir & 113226 & 63608 & G8III-IIIb (9) & 2.79 & 1.71 & 3.29 & 3.53 & 33.10 (10) & 2.44 & 83.43$\pm$59.80 & $\sim 700$ (13) \\
$\beta$ Oph & 161096 & 86742 & K2IIICN0.5 (9) & 2.75 (14) & 1.38 & 3.78 & 4.08 & 25.49 (10) & 2.72 & 43.62$\pm$36.25 & $\sim 2750$ (12) \\
$\lambda$ Sgr & 169916 & 90496 & K1IIIb (9) & 2.81 & 1.50 & 3.90 & 4.19 & 23.30 (10) & 1.68 & 86.78$\pm$35.39 & ... \\
$\alpha$ Ser & 140573 & 77070 & K2IIIbCN1 (9) & 2.63 (14) & 1.25 & 4.35 & 4.69 & 22.76 (10) & 2.65 & 17.63$\pm$53.77 & $\sim 2000$ (12) \\ \hline
\multicolumn{12}{c}{C-Rank Stars}\\\hline
$\epsilon$ Leo & 84441 & 47908 & G1IIIa (9) & 2.98 & 1.93 & 2.50 & 2.66 & 69.74 (10) & 3.45 & 155.17$\pm$119.69 & $\sim 210$ (12) \\
$\beta$ Dra & 159181 & 85670 & G2Ib-IIa (9) & 2.81 & 1.64 & 2.96 & 3.15 & 116.55 (2) & 3.15 & 496.41$\pm$154.89 & $\sim 65$ (8) \\
$\alpha$ Per & 20902 & 15863 & F5Ib (15) & 1.79 & ... & 3.16 & 3.35 & 155.28 (2) & 34.24** & ... & $\sim 41$ (8) \\
$\beta$ Crv & 109379 & 61359 & G5IIBa0.3 (16) & 2.64 & 1.59 & 3.13 & 3.34 & 45.44 (10) & 3.09 & 23.93$\pm$65.04 & $\sim 280$ (17) \\
$\alpha$ Aqr & 209750 & 109074 & G2Ib (9) & 2.94 & 1.76 & 3.19 & 3.39 & 202.22 (10) & 3.02 & 353.43$\pm$461.16 & $\sim 53$ (8) \\
$\delta$ CMa & 54605 & 34444 & F8Ia (7) & 1.84 & 1.00 & 3.23 & 3.42 & 492.61 (2) & 62.85** & ... & $\sim 12$ (8) \\
$\eta$ Dra & 148387 & 80331 & G8-IIIab (9) & 2.74 & 1.66 & 3.50 & 3.75 & 28.00 (10) & 3.71 & 87.03$\pm$88.47 & $\sim 650$ (12) \\
$\gamma^{2}$ Sgr & 165135 & 88635 & K0+III (9) & 2.99 & 1.75 & 3.52 & 3.79 & 30.99 (10) & 4.93 & 2628.63$\pm$167.72 & ... \\
\enddata

%% Include any \tablenotetext{key}{text}, \tablerefs{ref list},
%% or \tablecomments{text} between the \enddata and 
%% \end{deluxetable} commands

%% General table comment marker
\tablecomments{All $V$ and $I$ measurements are Johnson magnitudes sourced from \cite{ducati2022_johnson} except when indicated by a different reference. All $UDD_{V}$ and $LDD$ values are sourced from JSDC \citep{bourges2017}. RUWE is sourced from Gaia Data Release 3 \citep{gaiadr3} and tangential velocity anomaly ($dVt$) is taken from \cite{kervella2022} except when denoted by (**), for which both values are sourced from \cite{kervella2019}.}

%% General table references marker
\tablerefs{1. \cite{gray2003}, 2. \cite{vanleeuwen2007}, 3. \cite{rieutord2024}, 4. \cite{negueruela2024}, 5. \cite{pryzbilla2006}, 6. \cite{gray2006}, 7. \cite{gray2001}, 8. \cite{lyubimkov2010}, 9. \cite{keenan1989}, 10. \cite{gaia2020}, 11. \cite{lyubimkov2015}, 12. \cite{baines2018}, 13. \cite{howes2019}, 14. \cite{oja_2}, 15. \cite{skiff2014}, 16. \cite{lu1991}, 17. \cite{jofre2015}.}

\end{deluxetable*}

\vspace{-0.5in}
\section{Initial Vetting Observations, Data Reduction, and Analysis}\label{sec:observations}
To verify that these reference star candidates are usable for the Roman Coronagraph, any candidate with companions must be rejected. As many of the indirect methods of detection are often inconclusive and/or non-exhaustive, direct imaging and interferometry of these candidates is the most definitive approach to reference star vetting. Ideally, companion rejection limits should be measured down to \textit{TTR5} -- i.e., $10^{-7}$ contrast at Band 1 (575 nm) from $\sim 300-450$ mas -- to ensure the suitability of a reference star candidate. While no on-sky instrument is currently capable of achieving this level of performance, direct imaging observations can be used to set a companion rejection limit and to raise confidence in using a given reference star.

High-contrast instruments such as the Spectro-Polarimetric High-contrast Exoplanet REsearch instrument \citep[SPHERE;][]{beuzit2019}, the Subaru Coronagraphic Extreme Adaptive Optics system \citep[SCExAO;][]{lozi2018} with the Coronagraphic High Angular Resolution Imaging Spectrograph (CHARIS), and System for coronagraphy with High order Adaptive optics from R to K band \citep[SHARK-NIR;][]{farinato2022} achieve the deepest detection limits currently possible at 100--1000 mas separations. Additionally, powerful NIR interferometry instrumentation, such as the Michigan InfraRed Combiner-eXeter and the Michigan Young Star Imager at CHARA \citep[MIRC-X and MYSTIC;][]{anugu2020,monnier2018} and VLTI/GRAVITY \citep{gillessen2010}, is capable of directly detecting brighter companions at even smaller separations. However, observations of the entire reference star candidate list would take dozens of observing hours for any of these instruments. To address this need efficiently, the target list was first vetted with moderate-contrast imaging. High-contrast and interferometric instrumentation will be used in future observing campaigns to vet the remaining candidates more deeply.

\subsection{Direct Adaptive Optics (AO) Imaging with Hale/P3K+PHARO}\label{sec:PHAROobs}
We utilized the Palomar High Angular Resolution Observer \citep[PHARO;][]{hayward2001} installed at the Hale Telescope at Palomar Observatory to observe a subset of the CorGI-REx primary and reserve samples observable in the Northern sky in the 2024B and 2025A semesters (PIs: A. Greenbaum and C. Clark, respectively). The observations, summarized in Table \ref{tab:PHAROobs}, are efficient and require $<$15 minutes of integration time per target.

To avoid saturation, we mainly used the narrow $Br\gamma$ filter in combination with the narrow $H2$ filter in the grism wheel. If the risk of saturation was low, we used the $K_{\rm cont}$ filter instead. The data were collected in a 5-point quincunx (cross) dither pattern with each position separated by 5\arcsec\ from the previous position. At each position, 3 images were taken, offset from each other by 0.5\arcsec, for a total of 15 images per target. The sky background was estimated by median averaging all 15 images; each dithered imaged was then sky subtracted and flat-fielded. The reduced science frames were combined into a single combined image using intra-pixel interpolation that conserves flux, shifts the individual dithered frames by the appropriate fractional pixels, and median-coadds the frames. The final resolutions of the combined dithers were determined from the full-width half-maximum of the PSFs and were typically 0.1\arcsec.

The sensitivities of the final combined AO images were determined by injecting simulated sources azimuthally around the primary target every $20^\circ $ at separations of integer multiples of the central source's FWHM \citep{furlan2017}. The brightness of each injected source was scaled until standard aperture photometry detected the injected source with $5\sigma$ significance.

\begin{deluxetable*}{ccccccc}

%% Keep a portrait orientation

%% Over-ride the default font size
%% Use Default (12pt)

%% Use \tablewidth{?pt} to over-ride the default table width.
%% If you are unhappy with the default look at the end of the
%% *.log file to see what the default was set at before adjusting
%% this value.

%% This is the title of the table.
\tablecaption{Summary of AO imaging observations with Hale/P3K+PHARO.}\label{tab:PHAROobs}

%% This command over-rides LaTeX's natural table count
%% and replaces it with this number.  LaTeX will increment 
%% all other tables after this table based on this number
%\tablenum{3}

%% The \tablehead gives provides the column headers.  It
%% is currently set up so that the column labels are on the
%% top line and the units surrounded by ()s are in the 
%% bottom line.  You may add more header information by writing
%% another line between these lines. For each column that requries
%% extra information be sure to include a \colhead{text} command
%% and remember to end any extra lines with \\ and include the 
%% correct number of &s.
\tablehead{\colhead{Name} & \colhead{Date} & \colhead{Filter} & \colhead{$N \times t_{exp}$} & \colhead{$t_{int}$} & \colhead{5$\sigma$ Contrast} &\colhead{Approximate SpTy} \\ 
\colhead{} & \colhead{(UT)} & \colhead{} & \colhead{(s)} & \colhead{(s)} & \colhead{($\Delta mag$)} & \colhead{Sensitivity} } 

%% All data must appear between the \startdata and \enddata commands
\startdata
$\beta$ Cas & 2024 Aug 23 & $H2-Br\gamma$ & 15 $\times$ 60.9 & 913.5 & 7.34 & Mid-M \\
$\alpha$ Peg & 2024 Sep 22 & $H2-Br\gamma$ & 15 $\times$ 9.9 & 148.5 & 6.78 & Late-K \\
$\alpha$ Cyg & 2024 Sep 22 & $H2-Br\gamma$ & 15 $\times$ 9.9 & 148.5 & 7.31 & Early-B \\
$\eta$ Tau & 2025 Feb 10 & $H2-Br\gamma$ & 15 $\times$ 45.3 & 679.5 & 6.68 & Early-K \\
$\beta$ Tau & 2025 Feb 10 & $K_{cont}$ & 15 $\times$ 45.3 & 679.5 & 5.01 & Early-K \\
$\epsilon$ Gem & 2025 Feb 10 & $H2-Br\gamma$ & 15 $\times$ 45.3 & 679.5 & 4.34 & Early-B \\
$\beta$ CMi & 2025 Feb 10 & $K_{cont}$ & 15 $\times$ 1.4 & 21 & 4.76 & Mid-K \\
$\alpha$ Cep & 2025 Apr 6 & $H2-Br\gamma$ & 15 $\times$ 39.6 & 594 & 6.52 & Mid-M \\
$\epsilon$ Leo & 2025 Apr 6 & $H2-Br\gamma$ & 15 $\times$ 45.3 & 679.5 & 7.10 & Early-K \\
$\beta$ UMa & 2025 Apr 6 & $H2-Br\gamma$ & 15 $\times$ 2.8 & 42 & 6.84 & Early-M \\
$\delta$ Leo & 2025 Apr 6 & $H2-Br\gamma$ & 15 $\times$ 5.7 & 85.5 & 6.79 & Mid-M \\
$\beta$ Leo & 2025 Apr 6 & $H2-Br\gamma$ & 15 $\times$ 19.8 & 297 & 6.35 & Mid-M \\
$\epsilon$ UMa & 2025 Apr 6 & $H2-Br\gamma$ & 15 $\times$ 45.3 & 679.5 & 5.05 & Early-M \\
$\alpha^2$ CVn & 2025 Apr 6 & $H2-Br\gamma$ & 15 $\times$ 45.3 & 679.5 & 6.52 & Mid-M \\
$\eta$ UMa & 2025 Apr 6 & $H2-Br\gamma$ & 15 $\times$ 45.3 & 679.5 & 6.63 & Early-M \\
$\delta$ Crv & 2025 Apr 6 & $H2-Br\gamma$ & 15 $\times$ 45.3 & 679.5 & 6.21 & Early-M \\
$\beta$ Dra & 2025 Apr 6 & $H2-Br\gamma$ & 15 $\times$ 25.5 & 382.5 & 6.75 & Early-F \\
$\beta$ Lib & 2025 Apr 6 & $H2-Br\gamma$ & 15 $\times$ 45.3 & 679.5 & 6.12 & Early-M \\
$\zeta$ Oph & 2025 Apr 6 & $H2-Br\gamma$ & 15 $\times$ 45.3 & 679.5 & 6.51 & Late-G \\
$\zeta$ Aql & 2025 Apr 6 & $H2-Br\gamma$ & 15 $\times$ 45.3 & 679.5 & 6.22 & Early-M \\
$\gamma$ Cyg & 2025 Apr 6 & $H2-Br\gamma$ & 15 $\times$ 19.8 & 297 & 6.22 & Early-B \\
\enddata

%% Include any \tablenotetext{key}{text}, \tablerefs{ref list},
%% or \tablecomments{text} between the \enddata and 
%% \end{deluxetable} commands

%% No \tablecomments indicated
\tablecomments{Contrasts and approximate main sequence spectral type sensitivities were calculated at 0.\arcsec6 separation.}

%% No \tablerefs indicated

\end{deluxetable*}

\subsection{Speckle Interferometry with Gemini-North/`Alopeke and Gemini-South/Zorro}\label{sec:speckleobs}
In addition to AO imaging, we also performed speckle interferometry using the twin dual-channel imagers `Alopeke and Zorro \citep{scott2021} installed at the Gemini-North and Gemini-South observatories, respectively. Observations were performed over the 2024B and 2025A semesters (PIDs: GN-2024B-Q-213, GS-2024B-Q-204, GN-2025A-Q-113, GS-2025A-Q-111; PI: S. Wolff). The observations are summarized in Table \ref{tab:speckleobs}.

Speckle observations were performed at optical wavelengths at an elevation above 45$^{\circ}$ using narrow band filters. These constraints avoid atmospheric dispersion affecting the observations leading to false signals in the reduced data. Speckle imaging requires many thousands of short exposures (10–60 ms in length) to be obtained and processed. The short exposure length is necessary to ``freeze'' out the atmosphere in the observations and to obtain good speckle contrast. The large number of exposures is needed to build up sufficient S/N, especially at contrasts greater than five magnitudes down to the diffraction limit of the telescope. The observations are efficient; stars with $V$ magnitudes of 1-3 require only $\sim$5 min of observation time, during which 3000 to 5000 exposures are collected. Each reference star candidate observation is preceded or followed by an observation of a single PSF calibrator, also observed for a few minutes, to probe the atmospheric conditions experienced by the reference star at the time of observation. The observations in these programs were taken at 562 (blue) and 832 nm (red) simultaneously.

These exposures are reduced using Fourier analysis as described in \cite{horch2011} and \cite{howell2011}. Reduced data products include reconstructed images, robust 5$\sigma$ magnitude contrast limits in each bandpass, and, if a close companion is detected, information about its location and magnitude. The astrometric precision is typically $\pm$1 mas in angular separation and $\pm 1^{\circ}$ in position angle \citep{lester2021}. The photometric precision is typically $\pm$0.25 mag, increasing by $\sim 2 \times$ for very close or wider ($>0.85$\arcsec) pairs \citep[e.g.,][]{horch2011,howell2011}, where nonisoplanicity begins to dominate.

\begin{deluxetable*}{ccccccccccc}

%% Keep a portrait orientation

%% Over-ride the default font size
%% Use Default (12pt)

%% Use \tablewidth{?pt} to over-ride the default table width.
%% If you are unhappy with the default look at the end of the
%% *.log file to see what the default was set at before adjusting
%% this value.

%% This is the title of the table.
\tablecaption{Summary of speckle interferometry observations with Gemini-N/`Alopeke and Gemini-S/Zorro.}\label{tab:speckleobs}

%% This command over-rides LaTeX's natural table count
%% and replaces it with this number.  LaTeX will increment 
%% all other tables after this table based on this number
%\tablenum{1}

%% The \tablehead gives provides the column headers.  It
%% is currently set up so that the column labels are on the
%% top line and the units surrounded by ()s are in the 
%% bottom line.  You may add more header information by writing
%% another line between these lines. For each column that requries
%% extra information be sure to include a \colhead{text} command
%% and remember to end any extra lines with \\ and include the 
%% correct number of &s.
\tablehead{\colhead{Name} & \colhead{Date} & \colhead{Instrument} & \colhead{$N \times t_{exp}$} & \colhead{$t_{int}$} & \colhead{Calibrator} & \colhead{$N \times t_{exp}$} & \colhead{$t_{int}$} & 5$\sigma$ Blue Contrast & 5$\sigma$ Red Contrast & \colhead{Approximate SpTy} \\ 
\colhead{} & \colhead{(UT)} & \colhead{} & \colhead{(ms)} & \colhead{(s)} & \colhead{} & \colhead{(ms)} & \colhead{(s)} & ($\Delta mag$) & ($\Delta mag$) & \colhead{5$\sigma$ Sensitivity} } 

%% All data must appear between the \startdata and \enddata commands
\startdata
$\alpha$ Cep & 2024 Aug 14 & `Alopeke & $9000 \times 20$ & 180 & HR 8109 & $1000 \times 20$ & 20 & 6.31 & 7.30 & Late-K/Early-M \\
$\beta$ Cas & 2024 Aug 14 & `Alopeke & $9000 \times 20$ & 180 & HR 9079 & $1000 \times 20$ & 20 & 6.86 & 7.18 & Late-K/Early-M \\
$\delta$ Cas & 2024 Aug 15 & `Alopeke & $9000 \times 20$ & 180 & HR 481 & $1000 \times 20$ & 20 & 6.28 & 7.62 & Mid-Late K \\
$\alpha$ Peg & 2024 Aug 16 & `Alopeke & $9000 \times 20$ & 180 & HR 8784 & $1000 \times 20$ & 20 & 7.57 & 7.36 & Mid-K \\
$\alpha$ Gru & 2024 Sep 18 & Zorro & $9000 \times 14$ & 126 & HR 8440 & $1000 \times 14$ & 14 & 6.93 & 7.06 & Mid-K \\
$\alpha$ Hyi & 2024 Sep 18 & Zorro & $9000 \times 14$ & 126 & HR 571 & $1000 \times 14$ & 14 & 6.52 & 6.94 & Late-K/Early-M \\
$\eta$ Tau & 2024 Sep 27 & `Alopeke & $9000 \times 20$ & 180 & HR 1183 & $1000 \times 20$ & 20 & 7.73 & 7.61 & Mid-Late G \\
$\beta$ Eri & 2024 Nov 13 & Zorro & $12000 \times 14$ & 168 & HR 1673 & $1000 \times 14$ & 14 & 6.88 & 7.49 & Mid-Late K \\
$\alpha$ Lep & 2024 Nov 13 & Zorro & $12000 \times 14$ & 168 & HR 1742 & $1000 \times 14$ & 14 & 7.46 & 7.36 & Late-B/Early-A \\
$\beta$ CMa & 2024 Nov 13 & Zorro & $12000 \times 14$ & 168 & HR 2359 & $1000 \times 14$ & 14 & 7.71 & 7.88 & Late-F/Mid-G \\
$\alpha$ Col & 2024 Nov 14 & Zorro & $12000 \times 14$ & 168 & HR 1966 & $1000 \times 14$ & 14 & 7.23 & 7.23 & Late-G/Early-K \\
$\kappa$ Ori & 2024 Nov 14 & Zorro & $12000 \times 14$ & 168 & HR 1942 & $1000 \times 14$ & 14 & 7.07 & 7.59 & Early-Late F \\
$\epsilon$ Ori & 2024 Nov 14 & Zorro & $12000 \times 14$ & 168 & HR 1873 & $1000 \times 14$ & 14 & 6.67 & 7.08 & Late-B/Early-A \\
$\rho$ Pup & 2024 Nov 14 & Zorro & $12000 \times 14$ & 168 & HR 3123 & $1000 \times 14$ & 14 & 7.34 & 7.14 & Early-M \\
$\beta$ Tau & 2024 Nov 20 & `Alopeke & $12000 \times 14$ & 168 & HR 1768 & $1000 \times 14$ & 14 & 5.99 & 7.19 & Early-G/Early-K \\
$\alpha$ Cyg & 2024 Nov 21 & `Alopeke & $12000 \times 14$ & 168 & HR 4141 & $1000 \times 20$ & 20 & 6.61 & 7.20 & Late-B \\
$\gamma$ Ori & 2024 Nov 21 & `Alopeke & $9000 \times 14$ & 126 & HR 1807 & $1000 \times 20$ & 20 & 7.27 & 6.81 & Early-Mid G \\
$\beta$ CMi & 2024 Nov 21 & `Alopeke & $12000 \times 14$ & 168 & HR 2836 & $1000 \times 14$ & 14 & 7.73 & 7.41 & Mid-Late K \\
$\epsilon$ CMa & 2025 Jan 09 & Zorro & $18000 \times 10.3$ & 185.4 & HR 2637 & $1000 \times 10.3$ & 10.3 & 7.02 & 6.80 & Mid-F \\
$\eta$ CMa & 2025 Jan 10 & Zorro & $12000 \times 13.9$ & 166.8 & HR 2863 & $1000 \times 13.9$ & 13.9 & 7.72 & 7.44 & Early-A \\
$\zeta$ Pup & 2025 Jan 10 & Zorro & $12000 \times 13.9$ & 166.8 & HR 3162 & $1000 \times 13.9$ & 13.9 & 7.42 & 7.29 & Mid-A/Early-F \\
$\beta$ Car & 2025 Jan 10 & Zorro & $12000 \times 13.9$ & 166.8 & HR 3610 & $1000 \times 13.9$ & 13.9 & 6.58 & 7.49 & Late-G/Mid-K \\
$\iota$ Car & 2025 Jan 10 & Zorro & $12000 \times 13.9$ & 166.8 & HR 3673 & $1000 \times 13.9$ & 13.9 & 7.30 & 7.34 & Early-F \\
$\epsilon$ Gem & 2025 Feb 13 & `Alopeke & $12000 \times 20$ & 240 & HR 2346 & $1000 \times 20$ & 20 & 7.90 & 7.76 & Late-A/Late-F \\
$\epsilon$ Leo & 2025 Feb 14 & `Alopeke & $9000 \times 20$ & 180 & HR 3853 & $1000 \times 20$ & 20 & 7.52 & 7.41 & Early-K \\
$\alpha^2$ CVn & 2025 Feb 14 & `Alopeke & $9000 \times 20$ & 180 & HR 4816 & $1000 \times 20$ & 20 & 6.78 & 7.07 & Mid-K/Early-M \\
$\eta$ UMa & 2025 Feb 17 & `Alopeke & $12000 \times 15$ & 180 & HR 5148 & $1000 \times 15$ & 15 & 5.35 & 6.61 & Early-G/Mid-K \\
$\beta$ Lup & 2025 Mar 12 & Zorro & $12000 \times 15$ & 180 & HR 5580 & $1000 \times 15$ & 15 & 6.74 & 6.72 & Late-F/Mid-G \\
$\gamma$ TrA & 2025 Mar 13 & Zorro & $12000 \times 15$ & 180 & HR 5792 & $1000 \times 15$ & 15 & 6.45 & 6.81 & Late-G/Mid-K \\
$\beta$ TrA & 2025 Mar 13 & Zorro & $12000 \times 15$ & 180 & HR 5884 & $1000 \times 15$ & 15 & 6.83 & 6.31 & Early-M \\
$\beta$ Dra & 2025 Apr 14 & `Alopeke & $12000 \times 15$ & 180 & HR 6395 & $1000 \times 15$ & 15 & 5.65 & 7.18 & Early-Late F \\
$\zeta$ Aql & 2025 Apr 14 & `Alopeke & $9000 \times 20$ & 180 & HR 7173 & $1000 \times 20$ & 20 & 7.37 & 7.24 & Late-K/Early-M \\
$\zeta$ Oph & 2025 May 12 & Zorro & $9000 \times 20$ & 180 & HR 6189 & $1000 \times 20$ & 20 & 6.69 & 7.17 & Mid-F/Early-G \\
$\delta$ Cru & 2025 May 13 & Zorro & $9000 \times 20$ & 180 & HR 4592 & $1000 \times 20$ & 20 & 6.34 & 7.26 & Mid-F/Late-G \\
$\eta$ Cen & 2025 May 13 & Zorro & $9000 \times 20$ & 180 & HR 5431 & $1000 \times 20$ & 20 & 7.03 & 7.12 & Early-G/Early-K \\
$\gamma$ Cyg & 2025 Jun 11 & `Alopeke & $12000 \times 15$ & 180 & HR 7762 & $1000 \times 15$ & 15 & 5.76 & 6.80 & Mid-B \\
$\gamma$ Peg & 2025 Jun 12 & `Alopeke & $12000 \times 15$ & 180 & HR 86 & $1000 \times 15$ & 15 & 7.73 & 8.12 & Early-G/Early-K \\
$\epsilon$ Vir & 2025 Jun 14 & `Alopeke & $12000 \times 15$ & 180 & HR 4861 & $1000 \times 15$ & 15 & 7.23 & 7.53 & Mid-Late K \\
$\eta$ Dra & 2025 Jun 14 & `Alopeke & $12000 \times 15$ & 180 & HR 6198 & $1000 \times 15$ & 15 & 5.49 & 7.13 & Early-Late K \\
$\beta$ Oph & 2025 Jun 14 & `Alopeke & $12000 \times 15$ & 180 & HR 6633 & $1000 \times 15$ & 15 & 7.52 & 7.43 & Late-K \\
$\alpha$ Aqr & 2025 Jun 14 & `Alopeke & $12000 \times 15$ & 180 & HR 8495 & $1000 \times 15$ & 15 & 7.43 & 7.61 & Late-F \\
$\beta$ Aqr & 2025 Jul 03 & Zorro & $12000 \times 15$ & 180 & HR 8192 & $1000 \times 15$ & 15 & 5.71 & 6.68 & Late-A/Early-F \\
$\beta$ Lib & 2025 Jul 07 & Zorro & $12000 \times 15$ & 180 & HR 5669 & $1000 \times 20$ & 20 & 7.01 & 7.04 & Early-Mid K \\
\enddata

%% Include any \tablenotetext{key}{text}, \tablerefs{ref list},
%% or \tablecomments{text} between the \enddata and 
%% \end{deluxetable} commands

%% No \tablecomments indicated
\tablecomments{5$\sigma$ contrasts and approximate main sequence spectral type sensitivities were calculated at 0.\arcsec6 separation.}

%% No \tablerefs indicated

\end{deluxetable*}

\section{Results}\label{sec:results}
No new stellar companions were detected\footnote{Low significance companion signatures were identified in speckle interferometry observations of $\kappa$ Ori and $\gamma$ TrA, but were not significant enough to warrant a re-analysis of the data.} at $\geq 5\sigma$, although we do detect known stellar companions to $\alpha^2$ CVn ($\alpha^1$ CVn) and $\zeta$ Aql ($\zeta$ Aql B) at their expected positions in the PHARO-AO observations.

In the following sections, we describe the measured $5\sigma$ sensitivity curves from our observations in addition to extrapolated sensitivity curves at other wavelengths closer to the Roman Coronagraph bandpasses, assuming the observations were sensitive to bound main sequence companions. 
All reduced speckle interferometry and PHARO-AO imaging datasets and contrast curves are publicly accessible through the NASA Exoplanet Archive and Exoplanet Follow-up Observing Program \citep[ExoFOP;][]{akeson2013, christiansen2025, exofop5_doi}.

\subsection{PHARO Images and Sensitivity Curves} \label{sec:pharoimages}
\begin{figure*}
\includegraphics[width=\linewidth]{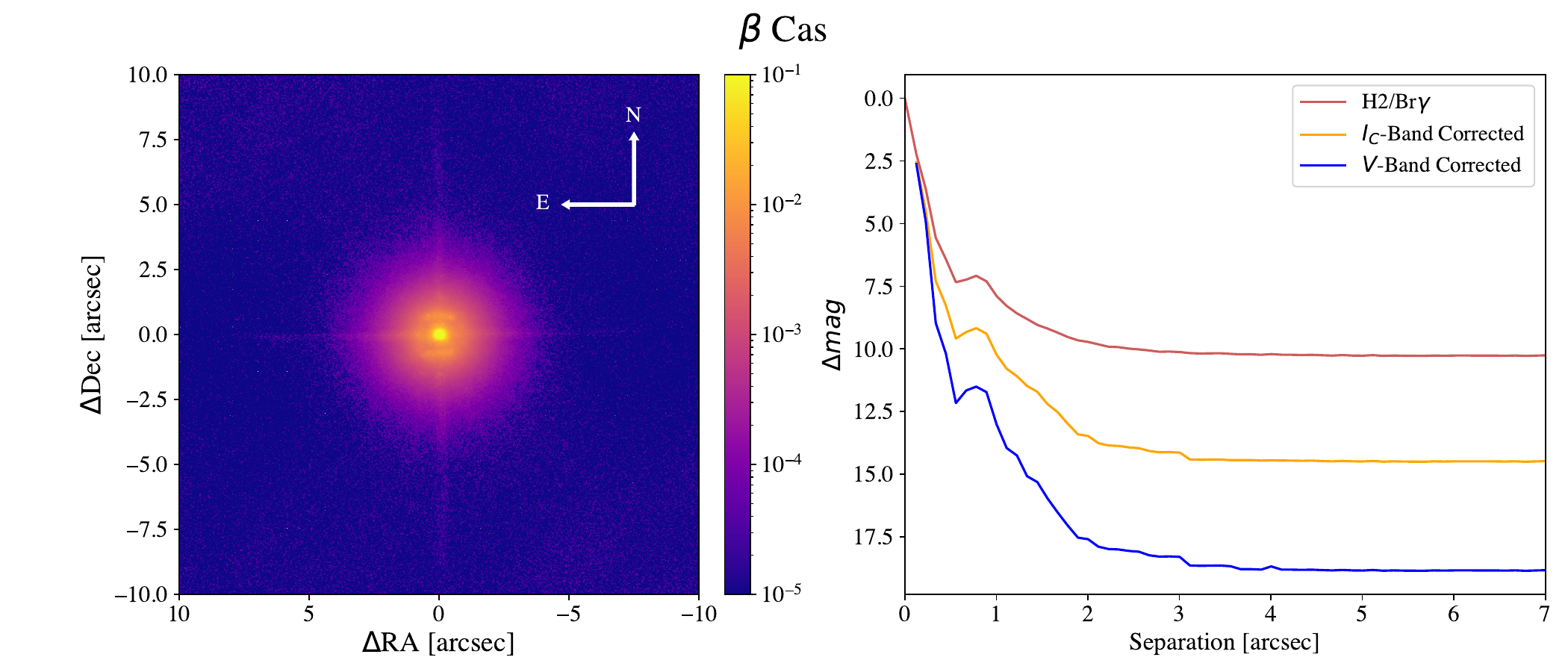}
\caption{PHARO $H2/Br\gamma$ image (\textit{left, contrast units}) and corresponding $5\sigma$ contrast curve measurement (\textit{right}) of $\beta$ Cas. Assuming bound main sequence stellar companion colors, we also show the $I_C$- and $V$-band corrected contrast curves derived from the $H2/Br\gamma$ contrasts and the system distance. The complete figure set (21 images) is available in the online journal.}\label{fig:PHAROimages}
\end{figure*}

The PHARO images and corresponding 5$\sigma$ contrast curves are shown in Figure \ref{fig:PHAROimages}. The median contrast achieved at 0.\arcsec6 was $\Delta mag \sim 6.5$. Given that the observations were taken at a wavelength close to the $Ks$-band, we can roughly predict the corresponding  $V$- and $I_C$-band contrasts (close to Roman Coronagraph Bands 1 and 4, respectively), assuming our observations were sensitive to bound companions not subject to extinction. To do this, we take the measured contrast curves in $\Delta mag$ and determine the closest spectral type match to a stellar companion at the same distance of the reference star candidate using the absolute magnitude values provided in \cite{pecaut2013}\footnote{Colors and absolute magnitudes in \cite{pecaut2013} for spectral types later than F0 are nominally main sequence values. For young systems, they are sufficiently similar to pre-main sequence values for the approximate spectral type sensitivities presented in this work.}. We then use the tabulated $V-K_S$ and $V-I_C$ colors to determine the corresponding $\Delta mag$ values for the $V$- and $I_C$-bands, respectively. For most reference star candidates within 100 pc, the PHARO observations are sensitive to stars as early as mid-G and as late as mid-M. Particularly for the M dwarfs, this corresponds to a much deeper contrast in $V$ than what was measured at H2+Br$\gamma$. For more distant stars, sensitivity starts to drop to early spectral types, where the corresponding companion sensitivities in $V$ and $I_C$ drop to similar to or even more shallow values.

The PHARO observations of $\alpha^2$ CVn presented in this work give a separation of known companion $\alpha^1$ CVn of $\sim$19\arcsec. With $V = 5.6$ ($\Delta mag\sim2.5$), this could introduce glint/off-axis speckles at a $10^{-8}$ contrast level into the HLC and SPC-WFOV dark holes. While we do not have wide-field observations of $\eta$ Dra in this work, the magnitude and separation of its known stellar companion $\eta$ Dra B ($\rho = $4.\arcsec93, $R = 8.29$), reveal that it could introduce glint/off-axis speckles at $\sim 10^{-8}$ and $\sim 10^{-9}$ levels in the SPC-WFOV and HLC FOVs respectively. Other reference star candidates ($\zeta$ Aql, $\epsilon$ CMa, $\delta$ Crv, $\beta$ Dra) also have relatively close stellar companions but are not expected to introduce incoherent light at detectable contrast levels.

\subsection{`Alopeke and Zorro Images and Sensitivity Curves} \label{sec:speckleimages}
\begin{figure*}
\includegraphics[width=\linewidth]{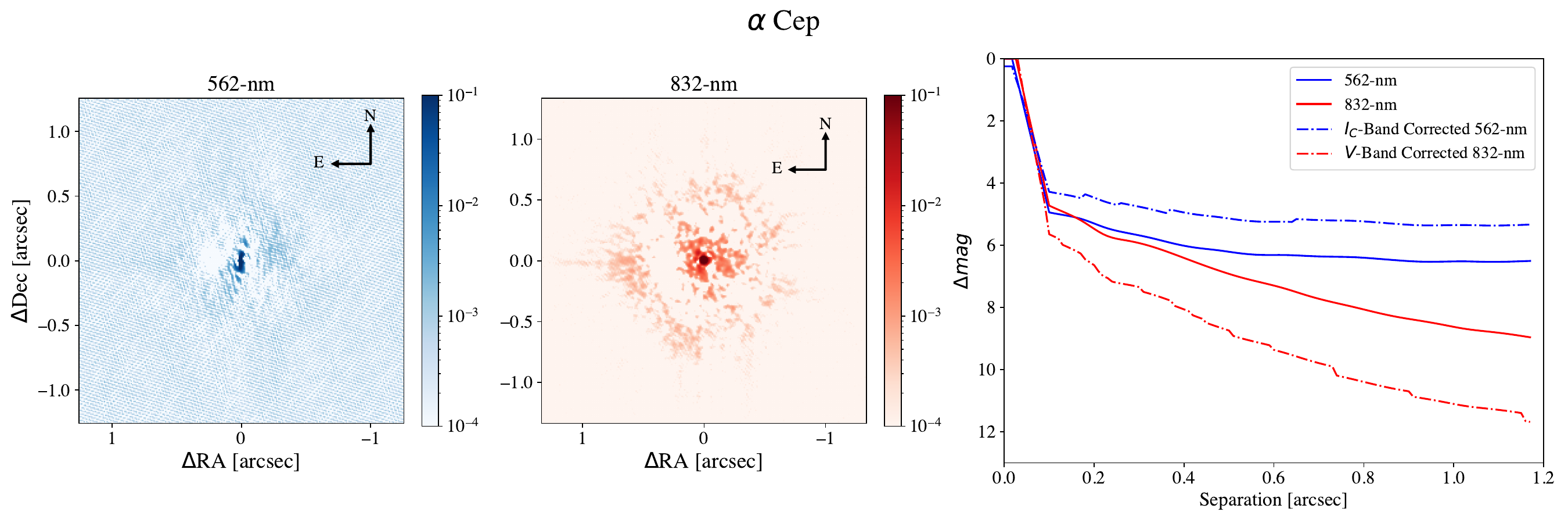}
\caption{`Alopeke 562- (\textit{left, contrast units}) and 832-nm images (\textit{center, contrast units}), and corresponding $5\sigma$ contrast curve measurements (\textit{right}) of $\alpha$ Cep. Assuming bound main sequence stellar companion colors, we also show the $I_C$- and $V$-band corrected contrast curves derived from the 562- and 832-nm contrasts, respectively, and the system distance. The complete figure set (43 images) is available in the online journal.}\label{fig:Speckleimages}
\end{figure*}

The `Alopeke and Zorro reconstructed images and corresponding 5$\sigma$ sensitivity curves are shown in Figure \ref{fig:Speckleimages}. The median contrast achieved at 0.\arcsec6 was $\Delta m_{562} \sim 6.9$ and $\Delta m_{832} \sim 7.2$. We compute color-corrected contrasts in the $V$-band for the 832-nm channel and in the $I_C$-band for the 562-nm channel by using the main sequence $V-I_C$ colors reported in \cite{pecaut2013}. Similar to the PHARO results, the speckle observations of nearby reference star candidates were sensitive to mid-to-late-type stellar companions (GKM) and only early-type companions around the most distant reference star candidates. While the speckle observations achieve similar overall contrasts to the AO observations, they achieve deeper contrasts at close separations that overlap with the FOV of the HLC observing mode.

\section{Ongoing Deeper Observations} \label{sec:futureobs}
Thus far, no literature-vetted (Tables \ref{tab:rex-corgi-sample-table} and \ref{tab:rex-corgi-reserve-table}) reference star candidates have been eliminated from consideration as a result of our low-contrast AO imaging and speckle interferometry observations. While the speckle interferometry and moderate-contrast AO imaging campaigns have been able to perform a preliminary vetting of a majority ($\sim85\%$) of our primary and reserve reference star samples, the achieved sensitivity limits are not sufficient for enabling full confidence in any reference star candidate.

The primary goal of CorGI-REx is to vet all primary reference star candidates. We will use higher-contrast AO and tighter-separation interferometric instrumentation across several observatories to obtain deeper and tighter-separation observations; the results of the CorGI-REx campaign and the investigation of archival direct imaging datasets will be reported in subsequent papers in the CorGI-REx series \citep{vega2026,schragal2026,brinjikji2026,lallemont2026} and will also be used to update current reference star rankings. CorGI-REx campaign prioritization will focus first on primary reference star candidates that lack any observations (archival or current) in a particular angular separation regime (2-200 mas interferometry and $>100$ mas high-contrast imaging). Prioritization within rankings is generally not considered. Some observations are queue-scheduled, where prioritization is generally not controlled. For classically-scheduled observations, there are often only a couple targets that are observable within a given observing window\footnote{High-contrast observations require observations around target transit passage to maximize the efficacy of post-processing ADI.}. Observations of reserve candidates are conducted at lowest priority if there are no other primary targets to observe.

\section{Discussion} \label{sec:discussion}

\subsection{Circumstellar Dust}
\label{sec:otherproperties}
The current reference star ranking scheme only considers evidence of bound companions that may contaminate the Roman Coronagraph dark hole, but any astrophysical source of incoherent light, such as bright circumstellar dust, may impact either/both HOWFSC and/or RDI efficiency and efficacy. Nearly a fifth of the primary reference star candidates have some combination of near-, mid-, and far-infrared (IR) excesses, likely from circumstellar dust emission \citep{su2006,chen2014,ertel2014,absil2013,rhee2007}, to varying degrees. Thus far, none of the debris disks around current reference stars with detected IR excesses have been detected in scattered light or resolved at any wavelength\footnote{Only one potential reserve reference star candidate ($\alpha$ PsA; Fomalhaut) with a warm debris disk was excluded because it was spatially resolved \citep{gaspar2023} with the Mid-Infrared Instrument (MIRI) on the James Webb Space Telescope (JWST).}. As a result, their resolved structure and sizes are unknown/poorly constrained, and estimations of their scattered light surface brightness and structure from thermal emission measurements are also likely to be poorly constrained \citep[e.g.,][]{hom2024,hom2025}, motivating their exclusion from reference star ranking consideration at this time. The presence of unresolved hot dust for reference stars with near-IR excesses has the potential to introduce effects similar to stellar leakage \citep[see also][]{douglas2019}, an effect already observed in JWST-MIRI coronagraphic imaging of systems with disk components interior to the inner working angle \citep{boccaletti2024}. Hot dust may also prove to be problematic for future observations of science target systems with unresolved exozodiacal dust \citep{ertel2025}. 

\subsection{Outlook for Roman Coronagraph Observation Scheduling} \label{sec:roman_obs}
The number of usable reference star candidates will have a significant impact on Roman Coronagraph science observations. The Coronagraph has a $\sim$2200-hour observing period after commissioning, and the ultimate scheduling of Roman Coronagraph observations must accommodate the scheduling of the WFI Core Community Surveys and in some cases cannot be interrupted \citep{roman2025}. Scheduling in general cannot be decided until a launch date is given, as the launch date impacts the ultimate location of Roman when commissioning and science operations commence.

\begin{figure*}
    \centering
    \includegraphics[width=\linewidth]{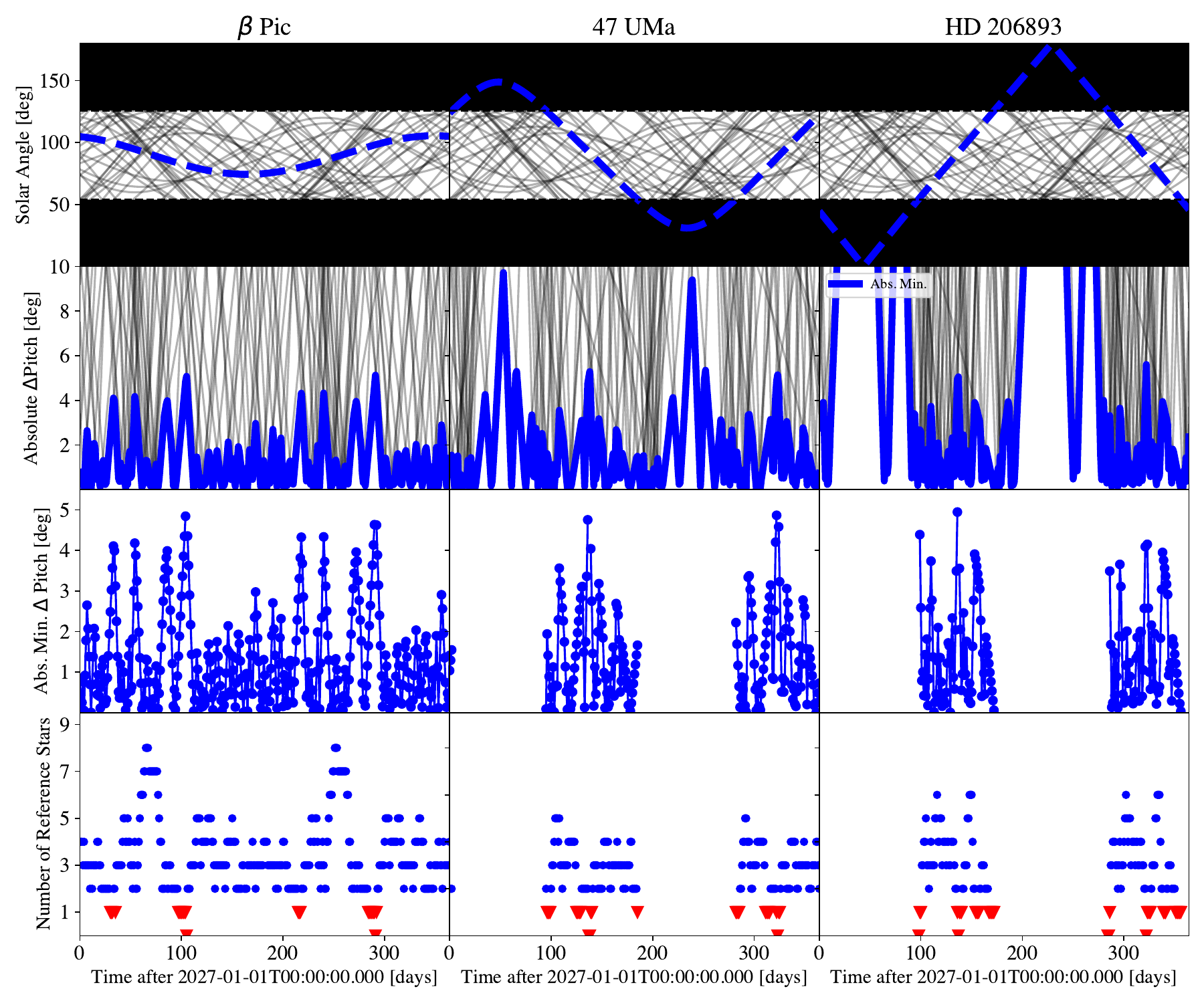}
    \caption{Outlook for scheduling Roman Coronagraph observations for three potential science targets (columns; $\beta$ Pic, 47 UMa, and HD 206893) assuming observations can commence January 1st, 2027 UT. The first row is the solar angle as a function of days of the year, with the dashed blue line corresponding to the solar angle of the science target and the gray lines corresponding to the solar angles of all considered reference stars. The shaded black regions are the solar keepout zones. A star is not observable any time of the year where a solar angle curve lies within either of these regions. The second row is the absolute value of $\Delta$pitch calculated between the science target and each primary reference star, and the blue curve is the absolute minimum $\Delta$pitch calculated among all primary reference star options on a given day. The third row combines information from the first two rows to determine the actual observation windows where both the science target and any primary reference star where $\Delta$pitch$\leq 5^{\circ}$ are observable. Finally, the fourth row shows the number of primary reference star options with $\Delta$pitch$\leq5^{\circ}$ that are observable with the science target on any given day. Red triangles indicate when $\leq1$ primary reference stars are available. If all primary reference star candidates are viable, scheduling is generally restricted only by science target observability.
    }
    \label{fig:scheduling_rankABC}
\end{figure*}

\begin{figure*}
    \centering
    \includegraphics[width=\linewidth]{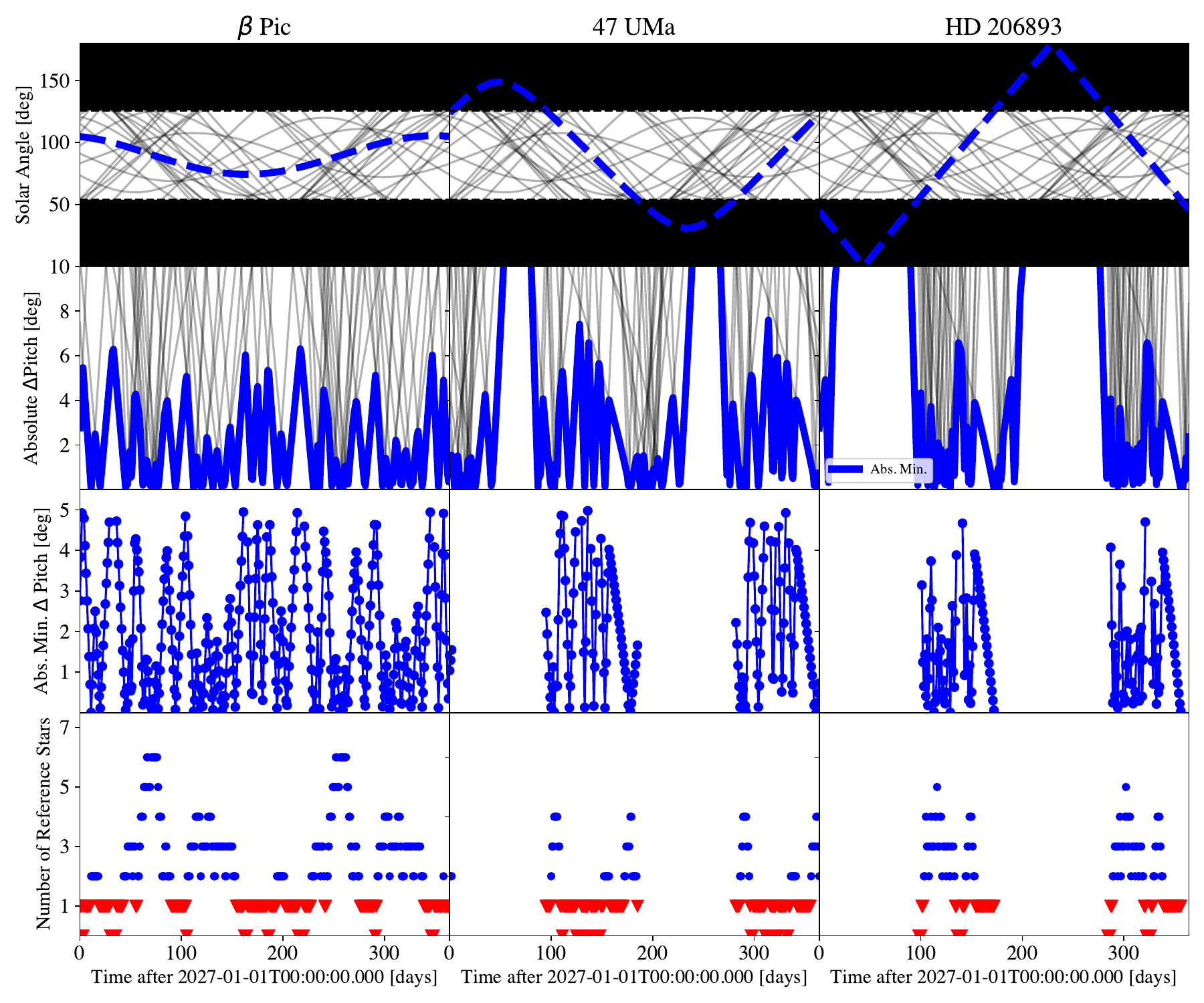}
    \caption{Same as Figure \ref{fig:scheduling_rankABC} but only considering Rank A and B primary reference stars.}
    \label{fig:scheduling_rankAB}
\end{figure*}

\begin{figure*}
    \centering
    \includegraphics[width=\linewidth]{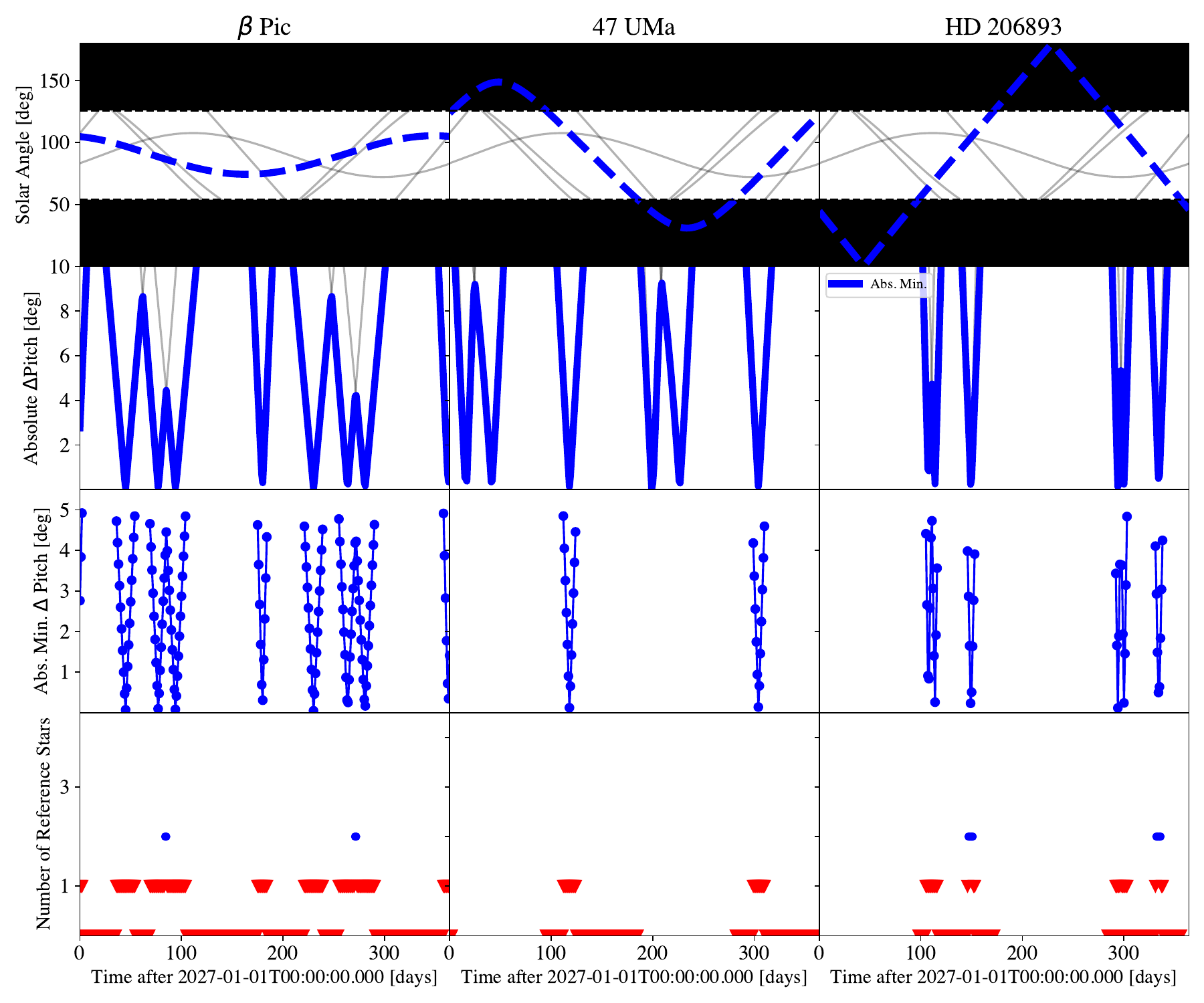}
    \caption{Same as Figure \ref{fig:scheduling_rankABC} but only considering Rank A primary reference stars.}
    \label{fig:scheduling_rankA}
\end{figure*}

The observability of targets in general is dependent on their locations in the sky relative to the positions of Roman and the Sun. The combination of reference and science star observability along with the $\Delta$pitch constraint limit the observability windows of Roman Coronagraph observations. To demonstrate the impact of scheduling and pitch angle constraints, we utilize the publicly available \texttt{roman-pointing} tool\footnote{\url{https://github.com/roman-corgi/roman_pointing/}} which queries the JPL Horizons Database for the ephemeris of the Roman Space Telescope to determine the observability windows of example science targets and the CorGI-REx primary sample. We consider three potential science targets, all of which are known to harbor exoplanets that the Roman Coronagraph may image in reflected light or self-luminous emission: (1) $\beta$ Pic \citep{lagrange2010,nowak2020}, (2) 47 UMa \citep{butler1996,fischer2002,gregory2010}, and (3) HD 206893 \citep{milli2017_hd206893}. These targets are intended to represent three distinct cases of general observability, with $\beta$ Pic, 47 UMa, and HD 206893 representative of high, medium, and low absolute ecliptic latitude targets, respectively. Assuming science observations start January 1, 2027, we calculate the solar angles, $\Delta$pitch angles, absolute minimum $\Delta$pitch angles, and number of available reference stars as a function of days of the year for all science and reference star combinations. We also calculate a scheduling efficiency, defined as the number of days the science target is observable (does not violate observatory pointing constraints relative to the Sun) and has an available reference star divided by the number of days the science target is observable by itself. In Figure \ref{fig:scheduling_rankABC}, we assume all Rank A, B, and C candidates are suitable. In Figure \ref{fig:scheduling_rankAB}, we assume only Ranks A and B, and in Figure \ref{fig:scheduling_rankA} we assume only Rank A. In Table \ref{tab:scheduling_days}, we show how many days during the year each science target is observable and has an observable reference star where $\Delta$pitch $\leq5^{\circ}$ depending on the reference star ranks being considered. For the most challenging observations (e.g., achieving $\sim10^{-9}$ contrast on $V>5$ stars), reference-science pairs must both be observable and have $\Delta$pitch$\leq5^{\circ}$ over spans of a few days.


\begin{deluxetable}{c|cccccc}

%% Keep a portrait orientation

%% Over-ride the default font size
%% Use Default (12pt)

%% Use \tablewidth{?pt} to over-ride the default table width.
%% If you are unhappy with the default look at the end of the
%% *.log file to see what the default was set at before adjusting
%% this value.

%% This is the title of the table.
\tablecaption{\textit{Left:} Number of days in a year each science target is observable with the Roman Coronagraph assuming there is an observable primary reference star where $\Delta$pitch $\leq5^{\circ}$. \textit{Right:} Corresponding scheduling efficiency. $\beta$ Pic, 47 UMa, and HD 206893 are observable within Solar keepout restrictions for 365, 177, and 148 days respectively.}\label{tab:scheduling_days}

%% This command over-rides LaTeX's natural table count
%% and replaces it with this number.  LaTeX will increment 
%% all other tables after this table based on this number
%\tablenum{1}

%% The \tablehead gives provides the column headers.  It
%% is currently set up so that the column labels are on the
%% top line and the units surrounded by ()s are in the 
%% bottom line.  You may add more header information by writing
%% another line between these lines. For each column that requries
%% extra information be sure to include a \colhead{text} command
%% and remember to end any extra lines with \\ and include the 
%% correct number of &s.
\tablehead{\colhead{Ranks} & \multicolumn2c{$\beta$ Pic} & \multicolumn2c{47 UMa} &\multicolumn2c{HD 206893} \\ 
\colhead{Considered} & \colhead{(days)} &\colhead{(\%)} & \colhead{(days)}&\colhead{(\%)} & \colhead{(days)}&\colhead{(\%)} } 

%% All data must appear between the \startdata and \enddata commands
\startdata
A, B, C & 363& 99 & 174&98 & 143&97 \\
A, B & 342& 88 & 155&88 & 132&89 \\
A & 130&36 & 25&14 & 39&26 \\
\enddata

%% Include any \tablenotetext{key}{text}, \tablerefs{ref list},
%% or \tablecomments{text} between the \enddata and 
%% \end{deluxetable} commands

%% General table comment marker
%\tablecomments{Comments}

%% General table references marker
%\tablerefs{References}

\end{deluxetable}

\begin{figure*}
    \centering
    \includegraphics[width=\linewidth]{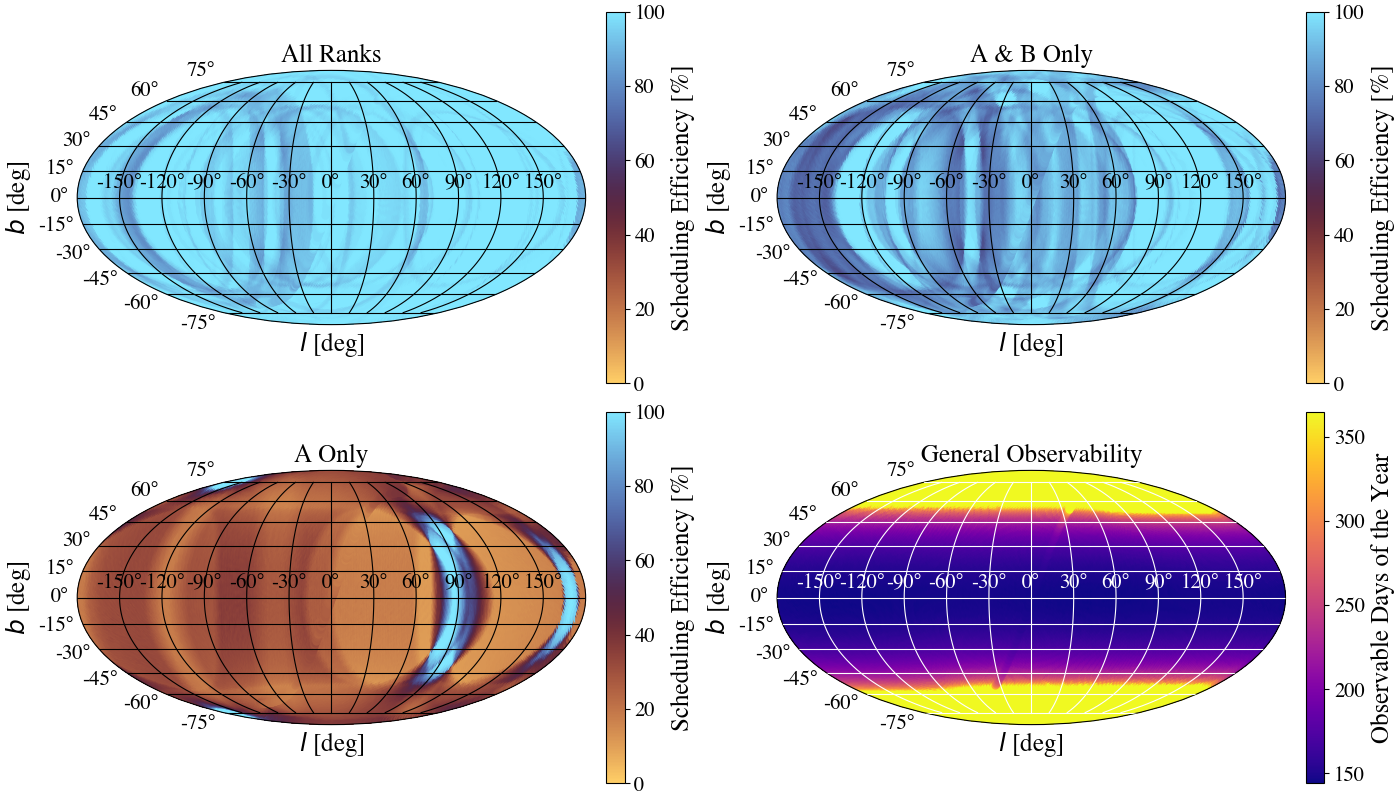}
    \caption{\textit{Top row and bottom left:} Scheduling efficiency heat maps in ecliptic coordinates for the all Ranks, Ranks A and B only, and Rank A only scenarios, for a $5^{\circ}$ $\Delta$pitch constraint. \textit{Bottom right:} Number of observable days in the year for the full sky. As shown in the top row, scheduling efficiency for the full sky is not significantly impacted by the number of available reference stars ($>20$). In the Rank A only scenario, the minimum and average scheduling efficiencies are 9.6\% and 30\% respectively, meaning that the scheduling of targets in most regions of the sky is extremely restrictive. High scheduling flexibility is maintained in similar ecliptic longitudes as Rank A stars $\beta$ CMa and $\kappa$ Ori (blue band $l \sim90^{\circ}$) and $\beta$ Leo (blue band $l \sim 170^{\circ}$) and in direct Solar angle proximity with Rank A star $\beta$ Car (blue regions around $l \sim -150^{\circ}$ and $b \sim \pm75^{\circ}$).}
    \label{fig:fullSkyRefCoverage}
\end{figure*}

We can also expand this analysis to the full sky by calculating $\Delta$pitch between each sky coordinate and all available primary reference stars. In Figure \ref{fig:fullSkyRefCoverage}, we show heat maps of the scheduling efficiencies for the three scenarios utilized in Figures \ref{fig:scheduling_rankABC}-\ref{fig:scheduling_rankA}.

Science targets in the CVZ (e.g., $\beta$ Pic) will always have more observability in general compared to targets outside of the CVZ, regardless of the number of usable reference stars. For science targets outside of the CVZ (e.g., 47 UMa and HD 206893), observability is only limited by the observability of the science target itself if a sufficient number of reference stars (Ranks A-C or Ranks A-B) are available, with scheduling efficiencies exceeding 85\%. For the full sky and all ranks, the minimum scheduling efficiency is 80\%, while the average is 97\%. For only Ranks A and B, the minimum scheduling efficiency is 63\% and the average is 90\%. As expected, if only Rank A stars are available, scheduling becomes extremely restrictive even for CVZ targets, with the average scheduling efficiency being 30\%. Despite this low average scheduling efficiency, the minimum scheduling efficiency in the Rank A scenario is only 9.6\%, meaning that no areas of the sky become completely unobservable if only Rank A reference stars can be used. In Table \ref{tab:scheduling_days}, it is also shown that HD 206893 has more observability than 47 UMa in the Rank A only scenario. This is because the reference star with the most observability with 47 UMa ($\beta$ UMa) is Rank B, and in the Rank A scenario the only reference star that could ever be used for 47 UMa is $\beta$ Car. Therefore, while Figures \ref{fig:scheduling_rankABC}-\ref{fig:scheduling_rankA} are meant to be representative of science targets at different ecliptic latitudes, the trend of increasing observability with increasing absolute ecliptic latitude will only hold if there is a sufficiently large selection of usable reference stars.

Thus far, no primary reference star candidates have been removed from initial consideration. If at least half of the reference star candidates are retained and found suitable, these scheduling restrictions are not expected to negatively impact Roman Coronagraph science efficiency.

\subsection{Implications for the Habitable Worlds Observatory}
HWO plans to survey a large sample of nearby FGK stars with the goal of directly imaging habitable zone exoEarths. This requires an optical/IR coronagraph capable of achieving $\sim10^{-10}$ contrast at separations less than 500 mas, which is only possible with the implementation of HOWFSC. 
If HWO adopts reference star HOWFSC (with or without RDI) with similar, or even more stringent, restrictions on reference star characteristics, the mission might use a sub-sample of the reference stars presented in this work. 
Additionally, many likely HWO science targets (FGK main sequence stars) have $B-V$ color indices ranging from 0.3-1.4, while OBA main sequence reference stars range from -0.33-0.27 \citep{pecaut2012,pecaut2013}. Color-mismatch is not an appreciable effect at the contrast levels reached by the Roman Coronagraph, but if color-matching between science and reference stars is introduced as a criterion for HWO to reach $10^{-10}$ contrast levels with RDI, only redder giants and subgiants (e.g., 12.5\% of our primary sample, 89\% of our reserve sample) would be potential color matches to FGK main sequence stars. To ensure that a potential HWO coronagraph instrument is not limited by these factors, the logical approach is to relax reference star criteria.

Reference star brightness is perhaps the most limiting criterion, as there are only 200 stars with $V<3$. The stellar diameter criterion cuts this number by more than half, and excluding systems with known companions further cuts this number by another factor of two, producing the 40-star reference star candidate sample described in this work. Thanks to its larger collecting area and anticipated improved core throughput \citep[e.g.,][]{feinberg2024,mennesson2024,redding2024}, HWO may be able to use fainter reference stars than Roman, but this remains to be quantified as HWO designs are further developed. If the magnitude threshold is increased to $V<5$, over 800 stars with $\mathrm{UDD}_V < 2$ mas could be considered (prior to vetting for companions).

The suitability criteria for HWO reference stars, considering only their utility for HOWFSC, may be loosened by allowing larger stellar angular diameters. The effect of stellar angular diameter on dark hole contrast is through ``incoherent" contrast, which does not affect the ability of HOWFSC to estimate coherent electric fields and produce DM solutions. If the stellar diameter criterion is relaxed to $\mathrm{UDD}_V < 10$ mas, an additional $\sim$50 stars with $V<3$ could be considered (some of which are already in our reserve sample). In the context of RDI, contrast degradation from stellar diameters are tied directly to coronagraph architecture; the HLC mode on Roman is very sensitive to low-order tip/tilt aberrations that could be caused by resolved stellar diameters. Other coronagraph architectures, such as the best-effort shaped-pupil coronagraphs or a vortex coronagraph \citep{mawet2010,mawet2013} are more robust to these effects and could potentially tolerate larger resolved stellar diameters if RDI is the baselined post-processing approach.

Relaxing the stellar multiplicity condition would allow many disqualified candidates to be used for HOWFSC and RDI. For HOWFSC, multi-star wavefront control \citep[MSWC;][]{thomas2015,sirbu2017} has been suggested as a potential method to suppress speckles from an off-axis companion. We note that MSWC with the Roman Coronagraph is not currently supported \citep{bendek2021,riggs2025}.

The pitch angle restriction is a major limitation to flexible and efficient scheduling for the Roman Coronagraph. As this is entirely the result of the impact of thermal variations while slewing between reference and science targets, HWO could consider designs that minimize these effects. If we repeat the exercise conducted in Figure \ref{fig:scheduling_rankA} but allow for $\Delta$pitch $\leq10^{\circ}$, observability is roughly doubled for $\beta$ Pic (218 days; 60\% scheduling efficiency), 47 UMa (52 days; 29\% scheduling efficiency), and HD 206893 (71 days; 48\% scheduling efficiency) despite only having 4 possible reference stars to choose from.

Other HOWFSC algorithms/approaches could be utilized to relax the frequency at which reference stars need to be observed. Dark zone maintenance \citep{redmond2024} has been demonstrated on the High-contrast imager for Complex Aperture Telescopes (HiCAT) testbed, and could relax the need for cyclic reference star observations.

Reference star characteristics may also motivate utilizing ADI post-processing as opposed to RDI. The dominant terms that make RDI less desirable for HWO are: 1) different broadband speckle morphologies created by color differences between stars with the same dark hole DM solution, 2) precision with which LOWFSC can position the differently colored stars on the same location of the FPM, and 3) the mismatch in stellar angular diameters between stars, with their resulting contrast contributions. If ADI is the preferred approach for post-processing, these reference star properties will not be as impactful to overall performance.

\section{Summary and Future Work} \label{sec:summary}
We have constructed an initial reference star candidate list that can be used for wavefront sensing and control techniques and reference differential imaging for the Roman Coronagraph and potentially an HWO coronagraph instrument. The candidates must be single, bright ($V<3$) stars with small ($\mathrm{UDD}_{V} < 2$ mas for the primary list, $\mathrm{UDD}_{V} < 5$ mas for the reserve list) angular diameters. The combination of these criteria limit the reference star candidate list to 40 possible options, with varying degrees of evidence that suggest stellar multiplicity.

Out of 95 bright stars with $\mathrm{UDD}_{V} < 5$ mas, 40 ($\sim42$\%) have confirmed multiplicity (although 4 are still suitable for HOWFSC). This is below the expected multiplicity rate of $\gtrsim50$\% for high-mass stars, although our sample is biased toward nearby systems and spans a wide range of spectral types and ages. This suggests that some current reference star candidates likely harbor problematic stellar companions that have yet to be discovered/confirmed and that our initial observations were not sensitive to. These undiscovered companions could be below the detection limit of archival investigations and our own observation campaigns or could also have been eclipsed by the primary star at the time of observation.

We conducted initial moderate-contrast AO imaging and speckle interferometry vetting observations to rule out any bright companions not reported in the literature and do not detect any previously unknown companions. The efficient preliminary vetting observations also allowed us to set a baseline companion rejection limit of $\Delta mag \sim 6$ for nearly all reference star candidates, improving the confidence in using them for Roman Coronagraph science operations. Deeper follow-up observations with powerful interferometric and high-contrast imaging instrumentation are underway and will be reported in later papers in this series.

Current reference star constraints are not expected to inhibit Roman Coronagraph observation efficiency as long as at least half of the primary reference star candidate list is suitable. Other reference star properties may have more subtle effects on achievable contrast on a case-by-case basis, and simulations of these effects would be valuable in determining their impact. If a potential coronagraph instrument onboard HWO has similar or more confining restraints, efficiency and science yield could be negatively impacted. Alternative wavefront sensing and control algorithms and observatory/instrument architectures may be able to relax constraints for future reference star selection.

\section*{Acknowledgements}
The authors would like to acknowledge the anonymous referee for their constructive, supportive, and expedient review.

The authors would also like to acknowledge the long list of astronomy community members, instrument principal investigators, and instrument team members who have volunteered time and resources into unofficial and official observations of CorGI-REx sample targets and observing proposal preparation. This list includes: John D. Monnier, Rachael Roettenbacher, Jeremy Jones, Denis Mourard, Becky Flores, Jean-Baptiste Le Bouquin, Jean-Philippe Berger, Manon Lallemont, Steve Ertel, Jacopo Farinato, Valentina D'Orazi, Dino Mesa, Fernando Pedichini, Simone Antoniucci, Olivier Guyon, Lucinda Lilley, Miles Lucas, Rebecca Zhang, Barnaby Norris, Julien Lozi, Thayne Currie, Connor Vancil, Jim Lyke, Eric Nielsen, Sloane Wiktorowicz, Jared Males, Joshua Liberman, Sebastiaan Haffert, Jennifer Patience, Jarron Leisenring, Logan Moore, Jun Hashimoto, Jennifer Power, Jacob Isbell, Jared Carlson, Eden McEwen, Katie Twitchell, Jay Kueny, Parker Johnson, Joseph Long, John Debes, and Katie Crotts.

This research has made use of the Washington Double Star Catalog maintained at the U.S. Naval Observatory. This research has made use of the JMMC Measured Stellar Diameters Catalogue\footnote{available at http://www.jmmc.fr/jmdc}. This research has made use of the Jean-Marie Mariotti Center JSDC catalogue, which involves the JMDC catalogue\footnote{JSDC available at http://www.jmmc.fr/jsdc}. This work made use of Astropy:\footnote{http://www.astropy.org} a community-developed core Python package and an ecosystem of tools and resources for astronomy \citep{astropy:2013, astropy:2018, astropy:2022}. This research has made use of the SIMBAD and VizieR databases, operated at CDS, Strasbourg, France.

Based on observations obtained at the international Gemini Observatory, a program of NSF NOIRLab, which is managed by the Association of Universities for Research in Astronomy (AURA) under a cooperative agreement with the U.S. National Science Foundation on behalf of the Gemini Observatory partnership: the U.S. National Science Foundation (United States), National Research Council (Canada), Agencia Nacional de Investigaci\'{o}n y Desarrollo (Chile), Ministerio de Ciencia, Tecnolog\'{i}a e Innovaci\'{o}n (Argentina), Minist\'{e}rio da Ci\^{e}ncia, Tecnologia, Inova\c{c}\~{o}es e Comunica\c{c}\~{o}es (Brazil), and Korea Astronomy and Space Science Institute (Republic of Korea). Some of the observations in the paper made use of the High-Resolution Imaging instruments `Alopeke and Zorro. `Alopeke and Zorro were funded by the NASA Exoplanet Exploration Program and built at the NASA Ames Research Center by Steve B. Howell, Nic Scott, Elliott P. Horch, and Emmett Quigley. `Alopeke and Zorro were mounted on the Gemini North and South telescopes of the international Gemini Observatory, a program of NSF NOIRLab, which is managed by the Association of Universities for Research in Astronomy (AURA) under a cooperative agreement with the U.S. National Science Foundation. on behalf of the Gemini partnership: the U.S. National Science Foundation (United States), National Research Council (Canada), Agencia Nacional de Investigaci\'{o}n y Desarrollo (Chile), Ministerio de Ciencia, Tecnolog\'{i}a e Innovaci\'{o}n (Argentina), Minist\'{e}rio da Ci\^{e}ncia, Tecnologia, Inova\c{c}\~{o}es e Comunica\c{c}\~{o}es (Brazil), and Korea Astronomy and Space Science Institute (Republic of Korea). This work has made use of data from the European Space Agency (ESA) mission {\it Gaia} (\url{https://www.cosmos.esa.int/gaia}), processed by the {\it Gaia} Data Processing and Analysis Consortium (DPAC, \url{https://www.cosmos.esa.int/web/gaia/dpac/consortium}). Funding for the DPAC has been provided by national institutions, in particular the institutions participating in the {\it Gaia} Multilateral Agreement.

J.~R.~H. is funded by NASA under award No. 80NSSC25K0364. S.~B.~H. acknowledges support from the NASA Exoplanets Program Office. D.~R.~C. acknowledges partial support from NASA Grant 18-2XRP18\_2-0007. This research has made use of the Exoplanet Follow-up Observation Program (ExoFOP; DOI: 10.26134/ExoFOP5) website, which is operated by the California Institute of Technology, under contract with the National Aeronautics and Space Administration under the Exoplanet Exploration Program. Based on observations obtained at the Hale Telescope, Palomar Observatory, as part of a collaborative agreement between the Caltech Optical Observatories and the Jet Propulsion Laboratory operated by Caltech for NASA. D.~S. is funded by NASA under award No. 80NSSC24K0216. Portions of this research were supported by funding from the Technology Research Initiative Fund (TRIF) of the Arizona Board of Regents and by generous anonymous philanthropic donations to the Steward Observatory of the College of Science at the University of Arizona. This research was carried out in part at the Jet Propulsion Laboratory, California Institute of Technology, under a contract with the National Aeronautics and Space Administration (80NM0018D0004). This material is based upon work supported by NASA under award Nos. 80NSSC24K0087, 80NSSC24K0097, and 80NSSC24K0217. This work was co-authored by employees of Caltech/IPAC under Contract No. 80GSFC21R0032 with the National Aeronautics and Space Administration.

The code used to create the figures in this manuscript are hosted publicly on Github (\url{https://github.com/jrhom1/rex-corgi-i}).

\software{numpy \citep{van2011numpy,harris2020array}, scipy \citep{jones2001scipy}, Astropy \citep{astropy:2013,astropy:2018,astropy:2022}, matplotlib \citep{matplotlib2007, matplotlib_v2.0.2}, iPython \citep{ipython2007}.}

\facilities{Gemini:Gillett, Gemini:South, Hale}

\appendix
\section{Summary of Rejected Reference Star Candidates} \label{sec:rejected_candidates}
In this section we list all reference star candidates rejected from our preliminary vetting analysis along with the rationale for rejection.
\subsection{Primary Sample Rejections}
\textit{$\beta^1$ Sco (HD 144217, HIP 78820):} The star has a 1.42 mas companion \citep{mason2010} and a $\sim$300 mas companion \citep{evans1978}.

\textit{$\gamma$ Cas (HD 5394, HIP 4427, Navi):} The star is a spectroscopic binary \citep{miroschnichenko2002} and has a bound 2.1\arcsec $\Delta mag \sim 9$ companion \citep{prasow-emond2024,burnham1889}.

\textit{$\alpha$ Lup (HD 129056, HIP 71860, Uridim):} The star has a 76 mas $\Delta K \sim 5.14$ companion \citep{gratton2023}.

\textit{$\epsilon$ Cen (HD 118716, HIP 66657):} The star has a 200 mas $\Delta mag \sim 2.5$ companion \citep{rizzuto2013}.

\textit{$\kappa$ Sco (HD 160578, HIP 86670):} The star has a 14.6 mas $\Delta mag \sim 4.2$ companion \citep{rizzuto2013}.

\textit{$\sigma$ Sgr (HD 175191, HIP 92855, Nunki):} The star is a spectroscopic binary \citep{chini2012,gullikson2013}, has a 8.6 mas $\Delta K \sim 0.1$ companion \citep{waisberg2025}, and is listed in BadCal.

\textit{$\delta$ Sco (HD 143275, HIP 78401, Dschubba):} The star is SBC9 1837 and has a 14 mas $\Delta mag \sim 3$ companion \citep{lebouquin2011}.

\textit{$\gamma$ Crv (HD 106625, HIP 59803, Gienah):} The star has a 1.\arcsec11 $\Delta K \sim 7$ companion \citep{janson2011}.

\textit{$\alpha$ Mus (HD 109668, HIP 61585):} The star has a 10 mas $\Delta mag \sim 2.8$ companion \citep{rizzuto2013}.

\textit{$\zeta$ Cen (HD 121263, HIP 68002, Leepwal):} The star is SBC9 793 and is a double-lined spectroscopic binary with predicted separation 1.2 mas \citep{hoffleit1991}.

\textit{$\beta$ Cru (HD 111123, HIP 62434, Mimosa):} The star has a 100 mas $\Delta mag \sim 6$ companion \citep{gratton2023} and is also known as SBC9 2446.

\textit{$\alpha^2$ Lib (HD 130841, HIP 72622, Zubenelgenubi):} The star is a spectroscopic binary \citep{fuhrmann2014} detected as a 19 mas $\Delta K \sim 0.3$ companion \citep{waisberg2023_alfLib}.

\textit{$\upsilon$ Sco (HD 158408, HIP 85696, Lesath):} \cite{hoffleit1991} categorizes the star as a spectroscopic binary. It also has a 7.8$\sigma$ proper motion anomaly \citep{kervella2019} and is listed in BadCal.

\textit{$\alpha$ CrB (HD 139006, HIP 76267, Alphecca):} The star is SBC9 850 \citep{ebbighausen1976,tomkin1986} with a predicted separation of $\sim$8 mas, $\Delta mag \sim 0.9$.

\textit{$\lambda$ Sco (HD 158926, HIP 85927, Shaula):} \cite{hoffleit1991} categorizes the star as a spectroscopic binary and \cite{tango2006} reports the star as a triple system.

\textit{$\alpha$ Oph (HD 159561, HIP 86032, Rasalhague):} The star has a 700 mas $\Delta mag \sim 2.9$ companion \citep{mccarthy1983}.

\textit{$\iota^1$ Sco (HD 161471, HIP 87073, Apollyon)):} The star is SBC9 986 and a spectroscopic binary in \cite{hoffleit1991}.

\textit{$\epsilon$ Sgr (HD 169022, HIP 90185, Kaus Australis):} The star has a 2.\arcsec4 $\Delta mag \sim 6.6$ companion \citep{golimowski1993}.

\textit{$\pi$ Sgr (HD 178524, HIP 94141, Albaldah):} The star is a triple system with one component of 100 mas separation, $\Delta mag \sim 0$, and another with 300 mas separation, $\Delta mag \sim 3$ \citep{finsen1965}.

\textit{$\alpha$ Pav (HD 193924, HIP 100751, Peacock):} The star is SBC9 1234 \citep{curtis1907} with a short period \citep{beavers1980}.

\textit{$\delta$ Cap (HD 207098, HIP 107556, Deneb Algedi):} The star is known as SBC9 1331 \citep{crump1921}.

\textit{$\delta$ Cyg (HD 186882, HIP 97165, Fawaris):} The star has a 2\arcsec $\Delta mag \sim 3.5$ companion \citep{struve1827}.

\textit{$\alpha$ Vir (HD 116658, HIP 65474, Spica):} The star is known as SBC9 766 with many components \citep{vogel1890}.

\textit{$\alpha$ And (HD 358, HIP 677, Alpheratz):} The star is known as SBC9 4 with separation 24 mas and $\Delta mag \sim 2$  \citep{pan1992}.

\textit{$\alpha$ Eri (HD 10144, HIP 7588, Achernar):} The star has a 300 mas $\Delta mag \sim 1.8$ companion \citep{kervella2008}.

\textit{$\beta$ Ari (HD 11636, HIP 8903, Sheratan):} The star is known as SBC9 98, a single-lined spectroscopic binary \citep{hoffleit1991} with 100 mas separation and $\Delta mag \sim 2.5$ \citep{pan1990}.

\textit{$\beta$ Per (HD 19356, HIP 14576, Algol):} The star is a tight triple system, also known as SBC9 157, with 2 mas, $\Delta mag \sim 3$ and  93 mas, $\Delta mag \sim 2$ separations and magnitudes respectively.

\textit{$\beta$ Aur (HD 40183, HIP 28360, Menkalinan):} The star is known as SBC9 366, and is eclipsing with a 0.08 AU semimajor axis and $\Delta mag \sim 0.2$ \citep{koechlin1983}.

\textit{$\theta$ Aur (HD 40312, HIP 28380, Mahasim):} The star has a 3.\arcsec7 $\Delta mag \sim 4.6$ companion \citep{roberts2011}.

\textit{$\gamma$ Gem (HD 47105, HIP 31681, Alhena):} The star is known as SBC9 410, with 0.\arcsec4 separation and $\Delta mag \sim 5.6$ \citep{sato1993}.

\textit{$\kappa$ Vel (HD 81188, HIP 45941, Markeb):} The star is known as SBC9 567 \citep{buscombe1960} and has Gaia DR3 RUWE $=30$.

\textit{$\gamma$ UMa (HD 103287, HIP 58001, Phecda):} The star is a 20.5 year orbit astrometric binary \citep{gontcharov2010} and proper motion derivative \citep{makarov2005,frankowski2007}.

\textit{$\delta$ Cen (HD 105435, HIP 59196):} The star has a 100 mas $\Delta mag \sim 2.9$ companion \citep{meilland2008}.

\textit{$\alpha$ Leo (HD 87901, HIP 49669, Regulus):} The star has a 15 mas $\Delta mag \sim 8$ WD companion \citep{gies2020}.

\subsection{Reserve Sample Rejections}

\textit{$\eta$ Boo (HD 121370, HIP 67927, Muphrid):} The star is known as SBC9 794 \citep{harper1910} with projected separation of 200 mas.

\textit{$\alpha$ PsA (HD 216956, HIP 113368, Fomalhaut):} The star has substantial resolved circumstellar dust at small and large separations \citep{gaspar2023,kalas2013}.

\textit{$\epsilon$ Aur (HD 31964, HIP 23416, Almaaz):} The star is known as SBC9 291, an 18 AU ``dark" eclipsing B-star companion \citep{hoard2010,kloppenborg2015}.

\section{Detailed Description of Reference Star Candidates}
\label{sec:rankings_detailed}
In this section, we provide a brief description of the rationale behind the B and C rankings of each primary and reserve reference star candidate along with general notes about all reference star candidates.

\subsection{Primary Candidates}
\textit{$\kappa$ Ori (HD 38771, HIP 27366, Saiph):} No previous literature has so far reported $\kappa$ Ori to possess any companions.

\textit{$\beta$ CMa (HD 44743, HIP 30324, Mirzam):} No previous literature has so far reported $\beta$ CMa to possess any companions, although \cite{kervella2019} reports a high RUWE.

\textit{$\beta$ Car (HD 80007, HIP 45238, Miaplacidus):} No previous literature has so far reported $\beta$ Car to possess any companions.

\textit{$\beta$ Leo (HD 102647, HIP 57632, Denebola):} While not having any significant companion detections, the star hosts a significant IR excess in the near, mid, and far-IR \citep{su2006,chen2014,absil2013}.

\textit{$\epsilon$ Ori (HD 37128, HIP 26311, Alnilam):} \cite{chini2012} reports this system as a double-lined spectroscopic binary. It is not mentioned in any other spectroscopic or astrometric binary catalogs.

\textit{$\delta$ Cas (HD 8538, HIP 6686, Ruchbah):} \cite{eggleton2008} reports the star as a possible eclipsing binary, but the eclipses have not been observed in recent years \citep{adelman2001}. \cite{zacharias2022} does not report it to be an astrometric binary. \cite{kervella2022} reports the system to have high RUWE but insignificant acceleration.

\textit{$\alpha$ Ara (HD 158427, HIP 85792):} The target is a Be star with a disk, and \cite{meilland2007} suggests the disk could be truncated by a companion. \cite{hoffleit1991} labels it as a spectroscopic binary, but astrometric binary evidence is inconclusive \citep{kervella2019}.

\textit{$\eta$ Cen (HD 127972, HIP 71352):} The star is classified as a spectroscopic binary in \cite{hoffleit1991} and a double-lined spectroscopic binary in \cite{chini2012}, but has no evidence of an astrometric perturbation \citep{zacharias2022}. \cite{kervella2019} reports a high RUWE.

\textit{$\rho$ Pup (HD 67523, HIP 39757, Tureis):} \cite{hoffleit1991} describes some evidence of the star having a long-period binary companion, but astrometric evidence is inconclusive \citep{kervella2022,kervella2019}. Finally, both \cite{chen2014} and \cite{rhee2007} reported far-IR excess detections of the order $\sim 10^{-5}$.

\textit{$\eta$ UMa (HD 120315, HIP 67301, Alkaid):} The target is listed as a questionable spectroscopic binary in \cite{hoffleit1991}.

\textit{$\gamma$ Ori (HD 35468, HIP 25336, Bellatrix):} \cite{hoffleit1991} states the star is a questionable spectroscopic binary, but no evidence is seen in \cite{chini2012}.

\textit{$\alpha$ Cyg (HD 197345, HIP 102098, Deneb):} \cite{hoffleit1991} states two possible configurations for spectroscopic binary companions in the system but is uncertain. It is also a pulsating star, which could result in radial velocity variations that could be confused for a spectroscopic binary signal.

\textit{$\beta$ Lup (HD 132058, HIP 73273):} \cite{hoffleit1991} lists the star as a spectroscopic binary, but does not give a source.

\textit{$\alpha$ Lep (HD 36673, HIP 25985, Arneb):} The target is a single star according to \cite{eggleton2008}, but \cite{zacharias2022} states it as an astrometric binary. \cite{kervella2019} reports a high RUWE but insignificant acceleration.

\textit{$\delta$ Leo (HD 97603, HIP 54872, Zosma):} This star is not described in many catalogs, but is labeled as an astrometric binary in \cite{zacharias2022}.

\textit{$\beta$ UMa (HD 95418, HIP 53910, Merak):} The star has an insignificant tangential acceleration and high RUWE \citep{kervella2022}, but is stated as an astrometric binary in \cite{zacharias2022}. The star also has detected mid- and far-IR excesses \citep{su2006,chen2014} of order $10^{-5}$.

\textit{$\eta$ CMa (HD 58350, HIP 35904, Aludra):} The star is not described in many catalogs, but is labeled as an astrometric binary in \cite{zacharias2022}.

\textit{$\alpha$ Cep (HD 203280, HIP 105199, Alderamin):} \cite{zacharias2022} states the star has an astrometric binary companion. \cite{absil2013} also detected a significant near-IR excess indicative of hot exozodiacal dust.

\textit{$\gamma$ TrA (HD 135382, HIP 74946):} The star has a low tangential acceleration \cite{kervella2022}, but has some evidence of harboring a lower mass stellar companion. \cite{waisberg2024_gamTrA} set an upper limit to direct detection of a possible companion with GRAVITY.

\textit{$\epsilon$ CMa (HD 52089, HIP 33579, Adhara):} The star does not have evidence of harboring a binary companion within the FOV of the Roman Coronagraph, but does have a 7.\arcsec5 $\Delta mag \sim 6$ companion that is not expected to introduce significant glint into any Roman Coronagraph observing mode FOV.

\textit{$\alpha$ Col (HD 37795, HIP 26634, Phact):} \cite{zacharias2022} labels it as an astrometric binary but \cite{kervella2022} reports an insignificant acceleration.

\textit{$\beta$ TrA (HD 141891, HIP 77952):} \cite{kervella2022} reports a marginal significance acceleration for the star.

\textit{$\alpha$ Gru (HD 209952, HIP 109268, Alnair):} \cite{chini2012} reports that the star has a constant radial velocity, but \cite{zacharias2022} reports the star as an astrometric binary.

\textit{$\beta$ CMi (HD 58715, HIP 36188, Gomeisa):} \cite{eggleton2008} reports to star to be a spectroscopic binary with a period of $\sim$200 days, with \cite{chini2012} reporting the system as a single-lined spectroscopic binary. \cite{kervella2019} reports a very high RUWE and low significance acceleration. \cite{harmanec2019} however refutes the spectroscopic binary classification and \cite{klement2024} did not detect any companions through CHARA interferometry observations.

\textit{$\alpha$ Peg (HD 218045, HIP 113963, Markab):} \cite{hoffleit1991} lists the star as a spectroscopic binary, but is unclear about the source. \cite{lee1910} suggests but does not directly state the star is a spectroscopic binary, but \cite{chini2012} reports the star having constant radial velocity.

\textit{$\zeta$ Pup (HD 66811, HIP 39429, Naos):} \cite{chini2012} describes the star as a double-lined spectroscopic binary, and \cite{zacharias2022} reports the star as an astrometric binary. This is also the reference star assumed in the Roman Coronagraph Observing Scenario 11 simulation suite\footnote{\url{https://roman.ipac.caltech.edu/page/coronagraph-public-images-html}}.

\textit{$\beta$ Cas (HD 432, HIP 746, Caph):} \cite{hoffleit1991} lists the star as a spectroscopic binary with a 27 day period, but the source is unclear. \cite{zacharias2022} describes the star as an astrometric binary. Both \cite{chen2014} and \cite{rhee2007} reported significant far-IR excess at a level of $\sim 10^{-4}$.

\textit{$\zeta$ Oph (HD 149757, HIP 81377):} \cite{chini2012} describes the star as a double-lined spectroscopic binary, and \cite{zacharias2022} reports the star as an astrometric binary. \cite{kervella2019} reports a very high acceleration and RUWE.

\textit{$\alpha$ Hyi (HD 12311, HIP 9236):} \cite{kervella2019} reports that the star has a high RUWE and marginally significant acceleration. \cite{zacharias2022} and \cite{malkov2012} also describe orbital solutions based on astrometric evidence.

\textit{$\eta$ Tau (HD 23630, HIP 17702, Alcyone):} WDSC lists the star as a lunar occultation binary, but interferometric studies \citep{hutter2021,klement2024} did not detect a companion. \cite{chini2012} also states that the system has a constant radial velocity. \cite{kervella2019} reports a low RUWE and low significance acceleration.

\textit{$\iota$ Car (HD 80404, HIP 45556, Aspidiske):} \cite{kervella2019} reports the system to have a $>5\sigma$ tangential velocity anomaly.

\textit{$\beta$ Tau (HD 35497, HIP 25428, Elnath):} WDSC lists the star as a lunar occultation binary of equal brightness and a 100 mas separation. \cite{adelman2006} lists the system as a spectroscopic binary but does not provide sufficient details.

\textit{$\delta$ Crv (HD 108767, HIP 60965, Algorab):} The star has a known 24\arcsec $\Delta mag \sim 5.5$ companion ($\delta$ Crv B) which is not expected to introduce significant off-axis glint into the dark hole. \cite{zacharias2022} labels the system as an astrometric binary, but it is ambiguous as to whether or not this originates from the known companion. The high RUWE of the system \citep{kervella2022} may also be attributed to the known companion. The star also has a significant near-IR excess, indicative of hot exozodiacal dust \citep{ertel2014}.

\textit{$\epsilon$ UMa (HD 112185, HIP 62956, Alioth):} WDSC lists the star as having a 100 mas companion with an unknown magnitude. \cite{kervella2019} and \cite{zacharias2022} provide stronger astrometric evidence. \cite{hoffleit1991} lists the star as a questionable spectroscopic binary.

\textit{$\beta$ Eri (HD 33111, HIP 23875, Cursa):} \cite{kervella2022} states the star has a high significance tangential acceleration. \cite{zacharias2022} also reports the system as an astrometric binary.

\textit{$\alpha^2$ CVn (HD 112413, HIP 63125, Cor Caroli):} The star has a known 19.\arcsec3 $\Delta mag \sim 2.6$ companion ($\alpha^1$ CVn) which could introduce contrast-degrading glint into the dark hole. The system also has a high RUWE which may be attributed to this companion \citep{kervella2022}.

\textit{$\beta$ Lib (HD 135742, HIP 74785, Zubeneschamali):} \cite{hoffleit1991} describes the star as a spectroscopic binary, but \cite{chini2012} states the star has a constant radial velocity. WDSC lists the star as having a disputed 2.\arcsec1 $\Delta mag \sim 9$ companion.

\textit{$\zeta$ Aql (HD 177724, HIP 93747, Okab):} The star has a known 7.\arcsec4 $\Delta mag \sim 9$ companion ($\zeta$ Aql B) that is not expected to introduce significant contrast-degrading glint into the dark hole. \cite{hoffleit1991} also describes the star as a spectroscopic binary. Finally, \cite{absil2013} reported a significant near-IR excess.

\textit{$\gamma$ Peg (HD 886, HIP 1067, Algenib):} \cite{makarov2005} describes the star as having a significant proper motion derivative, \cite{kervella2022} also reports a larger RUWE. \cite{chapellier2006} describes the star as a spectroscopic binary, but this assertion has been disputed given the pulsating nature of the star.

\subsection{Reserve Candidates}
\textit{$\alpha$ Aql (HD 187642, HIP 97649, Altair):} No previous literature has so far reported $\alpha$ Aql to possess any close companions. Several WDS components have sufficient separation and/or faintness to not cause any problems for wavefront sensing and control. \cite{kervella2022} reports a high RUWE but low significance acceleration.

\textit{$\beta$ Ori (HD 34085, HIP 24436, Rigel):} No previous literature has so far reported $\beta$ Ori to possess any close companions. A triple system located 9\arcsec away is not expected to introduce contrast-degrading glint.

\textit{$\beta$ Hyi (HD 2151, HIP 2021):} \cite{kervella2022} states that the system has a high RUWE (3.65) but insignificant acceleration. \cite{zacharias2022} also states that the system is an astrometric binary.

\textit{$\gamma$ Cyg (HD 194093, HIP 100453, Sadr):} \cite{kervella2019} states that the system has a very high RUWE (28.02) but there is otherwise no other evidence of a companion.

\textit{$\beta$ Aqr (HD 204867, HIP 106278, Sadalsuud):} \cite{kervella2022} states that the system has a high RUWE (3.08) but insignificant acceleration. \cite{zacharias2022} also states that the system is an astrometric binary.

\textit{$\epsilon$ Gem (HD 48329, HIP 32246, Mebsuta):} \cite{kervella2019} states that the system has a high RUWE (2.40) but other sources list the star as single.

\textit{$\epsilon$ Vir (HD 113226, HIP 63608, Vindemiatrix):} Both \cite{kervella2019} and \cite{kervella2022} state the system has high RUWE but low significance accelerations. \cite{zacharias2022} lists the star as an astrometric binary.

\textit{$\beta$ Oph (HD 161096, HIP 86742, Cebalrai):} Both \cite{kervella2019} and \cite{kervella2022} state the system has high RUWE but low significance accelerations. \cite{zacharias2022} lists the star as an astrometric binary.

\textit{$\lambda$ Sgr (HD 169916, HIP 90496, Kaus Borealis):} Both \cite{kervella2019} and \cite{kervella2022} state the system has high RUWE but low significance accelerations. \cite{zacharias2022} lists the star as an astrometric binary.

\textit{$\alpha$ Ser (HD 140573, HIP 77070, Unukalhai):} Both \cite{kervella2019} and \cite{kervella2022} state the system has a high RUWE but low significance accelerations. Other sources list the star as single.

\textit{$\epsilon$ Leo (HD 84441, HIP 47908, Algenubi):} Both \cite{kervella2019} and \cite{kervella2022} state the system has high RUWE but low significance accelerations. Other sources list the star as single.

\textit{$\beta$ Dra (HD 159181, HIP 85670, Rastaban):} The star has a known 4\arcsec.6 companion that is bound but too faint to introduce contrast-degrading glint. Both \cite{kervella2019} and \cite{kervella2022} state the system has high RUWE but marginally significant accelerations. \cite{zacharias2022} lists the star as an astrometric binary. It is unknown if these detections arise from the known companion.

\textit{$\alpha$ Per (HD 20902, HIP 15863, Mirfak):} \cite{kervella2022} reports an extremely high RUWE of 34.4.

\textit{$\beta$ Crv (HD 109379, HIP 61359, Kraz):} Both \cite{kervella2019} and \cite{kervella2022} state the system has high RUWE but marginally significant accelerations. Other sources state that the star is single.

\textit{$\alpha$ Aqr (HD 209750, HIP 109074, Sadalmelik):} Both \cite{kervella2019} and \cite{kervella2022} state the system has high RUWE but low significance accelerations. Other sources state that the star is single.

\textit{$\delta$ CMa (HD 54605, HIP 34444, Wezen):} \cite{kervella2022} reports an extremely high RUWE of 62.85.

\textit{$\eta$ Dra (HD 148387, HIP 80331, Athebyne):} Both \cite{kervella2019} and \cite{kervella2022} state the system has high RUWE but marginally significant accelerations. \cite{zacharias2022} reports that the system is not perturbed astrometrically, but \cite{eggleton2008} reports a 5.\arcsec1 $\Delta mag \sim 8.8$ companion ($\eta$ Dra B). This companion may introduce contrast-degrading glint into the dark hole.

\textit{$\gamma^2$ Sgr (HD 165135, HIP 88635, Alnasl):} \cite{kervella2019} and \cite{kervella2022} report inconsistent values of RUWE and tangential acceleration. \cite{zacharias2022} states that the star is an astrometric binary, and \cite{hoffleit1991} states that the star is a questionable spectroscopic binary.

\bibliographystyle{aasjournal}
\bibliography{export-bibtex}

\end{document}

%% file: rexcorgi_authors.txt
\author[0000-0001-9994-2142]{Justin Hom}
\affiliation{Steward Observatory and Department of Astronomy, University of Arizona, 933 N Cherry Avenue, Tucson AZ 85721}

\author[0000-0002-9977-8255]{Schuyler G. Wolff}
\affiliation{Steward Observatory and Department of Astronomy, University of Arizona, 933 N Cherry Avenue, Tucson AZ 85721}

\author[0000-0002-2361-5812]{Catherine A. Clark}
\affiliation{NASA Exoplanet Science Institute, IPAC, California Institute of Technology, Pasadena, CA 91125 USA}

\author[0000-0002-5741-3047]{David R. Ciardi}
\affiliation{NASA Exoplanet Science Institute, IPAC, California Institute of Technology, Pasadena, CA 91125 USA}

\author[0009-0002-9833-0667]{Sarah J. Deveny}
\affiliation{NASA Ames Research Center, Moffett Field, CA 94035, USA}
\affiliation{Bay Area Environmental Research Institute, Moffett Field, CA 94035, USA}

\author[0000-0002-2532-2853]{Steve B. Howell}
\affiliation{NASA Ames Research Center, Moffett Field, CA 94035, USA}

\author[0000-0002-7162-8036]{Alexandra Z. Greenbaum}
\affiliation{IPAC, Mail Code 100-22, Caltech, 1200 E. California Blvd., Pasadena, CA 91125, USA}

\author{Colin Littlefield}
\affiliation{NASA Ames Research Center, Moffett Field, CA 94035, USA}
\affiliation{Bay Area Environmental Research Institute, Moffett Field, CA 94035, USA}

\author[0000-0002-4989-6253]{Ramya M. Anche}
\affiliation{Steward Observatory and Department of Astronomy, University of Arizona, 933 N Cherry Avenue, Tucson AZ 85721}

\author[0000-0002-5407-2806]{Vanessa P. Bailey}
\affiliation{NASA Jet Propulsion Laboratory, California Institute of Technology, Pasadena, CA 91109, USA}

\author[0000-0003-1939-6351]{Wolfgang Brandner}
\affiliation{Max Planck Institute for Astronomy, K{\"o}nigstuhl 17, 69117 Heidelberg, Germany}

\author[0000-0003-4022-8598]{Ga{\"e}l Chauvin}
\affiliation{Max Planck Institute for Astronomy, K{\"o}nigstuhl 17, 69117 Heidelberg, Germany}
\affiliation{Laboratoire Lagrange, Universit{\'e} Cote d’Azur, CNRS, Observatoire de la Cote d’Azur, 06304 Nice, France}

\author[0000-0001-8627-0404]{Julien H. Girard}
\affiliation{Space Telescope Science Institute, 3700 San Martin Drive, Baltimore, MD 21218, USA}

\author[0009-0004-9449-137X]{Brian Kern}
\affiliation{NASA Jet Propulsion Laboratory, California Institute of Technology, Pasadena, CA 91109, USA}

\author[0000-0003-2008-1488]{Eric Mamajek}
\affiliation{NASA Jet Propulsion Laboratory, California Institute of Technology, Pasadena, CA 91109, USA}

\author[0000-0003-4205-4800
]{Bertrand Mennesson}
\affiliation{NASA Jet Propulsion Laboratory, California Institute of Technology, Pasadena, CA 91109, USA}

\author[0000-0002-8711-7206]{Dmitry Savransky}
\affiliation{Sibley School of Mechanical and Aerospace Engineering, Cornell University, Ithaca, NY 14853, USA}

\author[0000-0002-2805-7338]{Karl R. Stapelfeldt}
\affiliation{NASA Jet Propulsion Laboratory, California Institute of Technology, Pasadena, CA 91109, USA}

\author[0000-0003-4614-7035
]{Beth A. Biller}
\affiliation{Institute for Astronomy, University of Edinburgh, Royal Observatory, Edinburgh EH9 3HJ, UK}
\affiliation{Centre for Exoplanet Science , University of Edinburgh, Edinburgh, EH9 3FD, UK}

\author[0000-0002-0457-2941]{Marah Brinjikji}
\affiliation{Department of Physics, University of Notre Dame, South Bend, IN, USA}

\author[0000-0002-4677-9182
]{Masayuki Kuzuhara}
\affiliation{Astrobiology Center of NINS, 2-21-1, Osawa, Mitaka, Tokyo, 181-8588, Japan}
\affiliation{National Astronomical Observatory of Japan, 2-21-2, Osawa, Mitaka, Tokyo, 181-8588, Japan}

\author[0000-0001-6205-9233]{Maxwell A. Millar-Blanchaer}
\affiliation{Department of Physics, University of California, Santa Barbara, Santa Barbara, California, USA}

\author[0009-0002-9832-0004]{Toshiyuki Mizuki}
\affiliation{Astrobiology Center of NINS, 2-21-1, Osawa, Mitaka, Tokyo, 181-8588, Japan}
\affiliation{National Astronomical Observatory of Japan, 2-21-2, Osawa, Mitaka, Tokyo, 181-8588, Japan}

\author[0009-0002-8386-886X]{Nicholas T. Schragal}
\affiliation{Steward Observatory and Department of Astronomy, University of Arizona, 933 N Cherry Avenue, Tucson AZ 85721}

\author[0009-0002-5073-8671]{Macarena C. Vega-Pallauta}
\affiliation{Max Planck Institute for Astronomy, K{\"o}nigstuhl 17, 69117 Heidelberg, Germany}

\author[0000-0003-0774-6502]{Jason J. Wang}
\affiliation{Department of Physics and Astronomy, Northwestern University, 2145 Sheridan Road, Evanston, IL 60208-3112}
\affiliation{Center for Interdisciplinary Exploration and Research in Astrophysics, 1800 Sherman Ave,
Northwestern University, Evanston, IL 60201}

\author[0000-0002-4918-0247]{Robert J. De Rosa}
\affiliation{European Southern Observatory, Alonso de C\'{o}rdova 3107, Vitacura, Casilla 19001, Santiago, Chile}

\author[0000-0002-0813-4308]{Ewan S. Douglas}
\affiliation{Steward Observatory and Department of Astronomy, University of Arizona, 933 N Cherry Avenue, Tucson AZ 85721}

\author[0000-0003-1212-7538]{Bruce Macintosh}
\affiliation{Department of Astronomy and Astrophysics, University of California, Santa Cruz, Santa Cruz, CA 95064, USA}
\affiliation{University of California Observatories, 1156 High Street, Santa Cruz, CA 95064, USA}

\author[0000-0002-2696-2406]{Jingwen Zhang}
\affiliation{Department of Physics, University of California, Santa Barbara, Santa Barbara, California, USA}

\author{The Roman Coronagraph Community Participation Program}